\pdfoutput=1
\documentclass[11pt]{article}

\usepackage[preprint]{acl}

\usepackage{times}
\usepackage{latexsym}
\usepackage{amsmath,amssymb}
\usepackage{subcaption}
\usepackage{multirow}

\usepackage[T1]{fontenc}
\usepackage[utf8]{inputenc}

\usepackage{microtype}

\usepackage{inconsolata}

\usepackage{graphicx}
\usepackage{enumitem}
\usepackage{xcolor}
\usepackage{tcolorbox}

\usepackage{array}       
\usepackage{booktabs}    
\usepackage{colortbl}    
\usepackage{xcolor}      
\usepackage{graphicx}    

\title{\textsc{MultiConIR}: Towards Multi-Condition Information Retrieval}

\author{%
 \textbf{Xuan Lu\textsuperscript{1,2}},\;
 \textbf{Sifan Liu\textsuperscript{2}\thanks{Equal contribution}},\;
 \textbf{Bochao Yin\textsuperscript{2}\protect\footnotemark[\value{footnote}]},\;
 \textbf{Yongqi Li\textsuperscript{3}},\;
 \textbf{Xinghao Chen\textsuperscript{2,3}},\\
 \textbf{Hui Su\textsuperscript{4}},\;
 \textbf{Yaohui Jin\textsuperscript{1}},\;
 \textbf{Wenjun Zeng\textsuperscript{2}},\;
 \textbf{Xiaoyu Shen\textsuperscript{2}\thanks{Corresponding author}}\\[4pt]
 \textsuperscript{1}Shanghai Jiao Tong University\\
 \textsuperscript{2}Ningbo Key Laboratory of Spatial Intelligence and Digital Derivative, Institute of Digital Twin, EIT\\
 \textsuperscript{3}Department of Computing, The Hong Kong Polytechnic University\\
 \textsuperscript{4}Meituan Inc.\\
 \texttt{lux1997@sjtu.edu.cn\quad xyshen@eitech.edu.cn}
}

\begin{document}
\maketitle
\begin{abstract}
Multi-condition information retrieval (IR) presents a significant, yet underexplored challenge for existing systems. This paper introduces \textbf{\textsc{MultiConIR}}, a benchmark specifically designed to evaluate retrieval and reranking models under nuanced multi-condition query scenarios across five diverse domains. We systematically assess model capabilities through three critical tasks: complexity robustness, relevance monotonicity, and query format sensitivity. Our extensive experiments on 15 models reveal a critical vulnerability: most retrievers and rerankers exhibit severe performance degradation as query complexity increases. Key deficiencies include widespread failure to maintain relevance monotonicity, and high sensitivity to query style and condition placement. The superior performance of GPT-4o reveals the performance gap between IR systems and advanced LLM for handling sophisticated natural language queries. Furthermore, this work delves into the factors contributing to reranker performance deterioration and examines how condition positioning within queries affects similarity assessment, providing crucial insights for advancing IR systems towards complex search scenarios. 
The code and datasets are available at \url{https://github.com/EIT-NLP/MultiConIR}
\end{abstract}

\section{Introduction}

Information retrieval (IR) is critical for helping users find relevant information across various domains.  Traditionally, IR systems retrieve documents by matching queries based on lexical similarity, such as BM25~\citep{carpineto2012survey, ponte2017language}, or semantic similarity using dense vector representations~\citep{karpukhin-etal-2020-dense,zhan2021optimizing}. 
Though highly effective for queries with straightforward query-document relationships~\citep{su2024bright}, they often fail to fully capture nuanced intent as user needs become more complex~\citep{zhu2023large,su2024bright,zhu2025collaborative}. 

\begin{figure}
    \centering
    \includegraphics[width=1\linewidth]{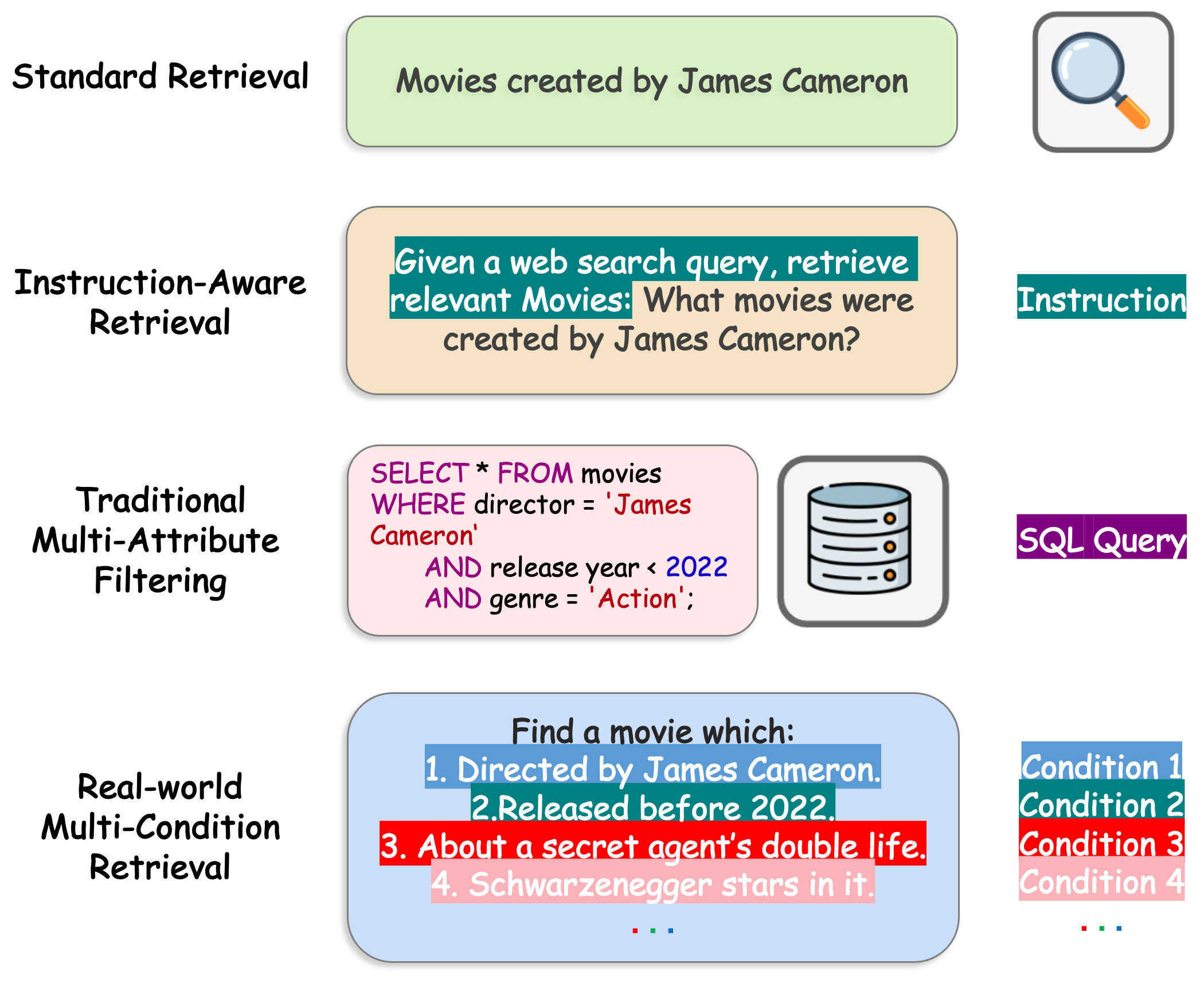}
    \caption{\small From single-condition to multi-condition retrieval. \small Standard and instruction-aware retrieval address single-condition queries. SQL-based filtering is restricted to predefined attributes within structured data. Real-world multi-condition retrieval enables the formulation of multiple, often semantic, conditions using natural language query.
}
    \label{fig:introduction}
\end{figure}

A significant challenge arises when users specify multiple requirements simultaneously, as illustrated in Fig.~\ref{fig:introduction}. Whether searching for a movie with specific attributes or selecting a product that meets various criteria, multi-condition search has become an integral part of modern information-seeking behavior.
Traditional IR systems handle such scenarios using structured filtering, such as SQL-based queries that retrieve information from backend databases based on predefined conditions. However, this approach is inherently rigid and limited, as it relies on explicitly defined attributes and lacks the flexibility to accommodate evolving or diverse user preferences. As a result, it struggles to support nuanced or semantic-level queries that go beyond structured data filtering.

The advent of Large Language Models (LLMs) has enhanced IR by introducing instruction-following capabilities~\citep{asai2023task,weller2024followir,oh2024instructir,liu2024rag}. This approach augments standard queries with explicit instructions, which serve as additional constraints to refine search results, as shown in Fig.\ref{fig:introduction}.
Despite these advancements, existing evaluation benchmarks remain predominantly focused on single-condition queries and binary relevance assessments—classifying documents as either relevant or irrelevant~\citep{nguyen2016ms,kwiatkowski2019natural,muennighoff2022mteb}—thus overlooking the nuanced challenges of multi-condition queries, where relevance depends on the degree to which multiple conditions are satisfied.

An ideal multi-condition retrieval system should exhibit the following properties:
(1) \textbf{Complexity Robustness:} The system should maintain high performance regardless of query complexity (i.e., the number of conditions specified);
(2) \textbf{Relevance Monotonicity:} The relevance scores should scale monotonically with the number of matched conditions; for example, a document matching all $n$ conditions should be ranked higher than one matching $n-1$;
(3) \textbf{Format Invariance:} The system should yield consistent results regardless of the query format, whether presented as a structured list or as free-form natural language.

Existing benchmarks do not offer a structured framework for evaluating multi-condition retrieval along these dimensions. To address this gap, we introduce \textbf{\textsc{MultiConIR}}—a benchmark designed to comprehensively evaluate multi-condition retrieval systems. Through systematic experiments on 15 state-of-the-art models (including dense retrievers, cross-encoders, and LLM-based agents), we uncover several critical insights:

\begin{itemize}
\item \textbf{Multi-Condition Struggle:} Retrievers and Rerankers struggle with multi-condition retrieval, showing performance decline as query conditions increase, difficulty with relevance monotonicity, and sensitivity to query style variations.  

\item \textbf{Retrievers and Rerankers Differ:} Rerankers excel with single-condition queries but fail under multiple conditions. Retrievers demonstrate greater robustness. GritLM demonstrates the best robustness among retrievers.

\item \textbf{Position Impacts Model Focus:} Dense retriever pooling strategies emphasize different condition positions, mean pooling focuses on initial positions, while \texttt{<EOS>} pooling emphasizes final positions.  Rerankers exhibit non-uniform attention across positions and greater sensitivity to document length variations. 

% Rerankers' token-level interactions benefit single-condition tasks but hinder multi-condition retrieval.
% \item \textbf{Relevance Monotonicity Struggle:} Models fail to maintain a consistent ranking hierarchy.
% % Dense retrievers tend to focus more on “exact match” and “complete mismatch”, likely due to their training paradigm.
% \item \textbf{Query Format Sensitivity:} Models are susceptible to query phrasing, with rerankers exhibiting a flip rate exceeding 20\%.
% \item \textbf{Rerankers' Sensitivity to Complex Input:} Rerankers demonstrate notable susceptibility to input complexification, stemming from an increase in query conditions, variations in query style, or greater document length.
% \item \textbf{Position Impact on Focus:} Different pooling methods focus on different condition positions. Rerankers exhibit a non-uniform distribution of attention across different positions.
% \item \textbf{Pooling Impact on Focus:} Different embedding pooling strategies influence the model’s focus on query condition positions. Mean pooling focuses on early query conditions, \texttt{<EOS>} pooling biases toward later conditions, while latent attention layer pooling distributes focus more evenly.
% \item \textbf{GritLM’s Advantage:} GritLM outperforms other models, potentially due to its hybrid attention mechanism, which integrates both bidirectional and causal attention to improve the understanding of complex queries.
\end{itemize}

By quantifying these gaps, our work reveals key deficiencies in the ability of current IR systems to understand multi-condition intent, laying the groundwork for advancing IR toward human-like reasoning in complex search scenarios.

\section{Related Works}
\paragraph{Retriever: From Sparse To Dense}
Traditional sparse retrieval methods are based on BM25 \citep{robertson2009bm25}, 
TF-IDF \citep{ramos2003using}, etc., rely on keyword matching and statistical weighting to evaluate relevance, which suffers from the well-known issue of lexical gap \citep{berger2000bridging}, 
restricting their ability to effectively capture semantic relationships \citep{luan2021sparse, Nian2024WRAG}.
Dense retrieval addresses this limitation by encoding both queries and documents as embeddings within a joint latent space, where the semantic relationship is captured through the similarity scores between their embeddings \citep{li2023makinglargelanguagemodels}. Pre-trained language models like BERT \citep{devlin2019bert} and T5 \citep{raffel2020exploring}, are widely used as backbone encoders for dense retrieval \citep{li2023towards,sturua2024jina,bge_embedding}.
Recent advancements have shown that LLMs offer significant potential as backbone encoders for dense retrieval \citep{wang-etal-2024-improving-text, weller2024promptriever, behnamghader2024llm2veclargelanguagemodels}. 
For instance, Repllama~\citep{ma2023finetuning} fine-tuned Llama-2 to serve as dense retrievers. 
GritLM~\citep{muennighoff2024generative} unified text embedding and generation within a single LLM. LLM2Vec~\citep{behnamghader2024llm2veclargelanguagemodels} introduced an unsupervised approach for transforming decoder-only LLMs into dense retrievers.

\paragraph{Benchmarks In Complex Retrieval Tasks}

Existing datasets for information retrieval, such as MS MARCO \citep{nguyen2016ms}, Natural Questions \citep{kwiatkowski2019natural}, and MTEB \citep{muennighoff2022mteb}, primarily focus on queries sourced from search engines. The relationships between queries and documents are typically simple and direct \citep{su2024bright}.
Recent studies have expanded retrieval benchmarks to address more complex scenarios. BIRCO \citep{wang2024bircobenchmarkinformationretrieval} introduces a benchmark designed to evaluate information retrieval tasks with complex objectives. Instruction-based datasets~\citep{weller2024followir, qin2024infobench, oh2024instructir}, for instance, evaluate the instruction-following capabilities of retrieval models by embedding explicit instructions within queries to better represent users’ retrieval intents. Furthermore, some works have assessed retrieval models’ abilities to handle logical reasoning tasks, including Boolean logic~\cite{mai2024setbert, zhang2024boolquestionsdoesdenseretrieval}, negation~\cite{zhang2024excluir, weller2024nevirnegationneuralinformation}, and multi-hop reasoning~\cite{su2024bright,liu2025qfft}. These efforts mark significant progress in increasing query complexity. However, while research in the generative modeling domain has explored the ability of LLMs to handle multi-constraint instructions \citep{he2024complex, ferraz2024llm, zhang2024iopo}, studies on retrieval models in multi-condition scenarios remain sparse.

\section{\textsc{MultiConIR}}
We introduce \textsc{MultiConIR}, a benchmark designed to evaluate the capacity of retrieval models to process multi-condition queries. Formally, given a query \( q_k \) composed of \( k \) conditions \( C = \{c_1, c_2, \dots, c_k\} \) with \( k \in \{1, \dots, 10\} \), we construct a structured retrieval setup consisting of:

(1) \textbf{Two query formulations}, denoted as \( q_k^{\text{inst}} \) and \( q_k^{\text{desc}} \), where \( q_k^{\text{inst}} \) corresponds to a structured instruction-style query, formally expressed as a tuple \( q_k^{\text{inst}} = \langle f, C \rangle \) where \( f \) is an explicit function describing retrieval constraints, and \( q_k^{\text{desc}} \) is a natural language descriptive-style query, represented as an unstructured sequence about the same set \( C \),

(2) A \textbf{positive document} \( d^+ \) that satisfies all \( k \) conditions, i.e., \( d^+ \models C \),  

(3) A sequence of \textbf{hard negative (HN) documents} \( \{d_0, d_1, \dots, d_{k-1}\} \), where each \( d_j \) satisfies exactly \( j \) out of \( k \) conditions, formally expressed as \( d_j \models \{c_1, \dots, c_j\} \) and \( d_j \not\models \{c_{j+1}, \dots, c_k\} \).

This controlled design enables a principled evaluation of multi-condition retrieval along three fundamental axes: (1)  \textbf{Complexity Robustness:} The model's retrieval effectiveness as \( k \) increases, measured by its ability to distinguish \( d^+ \) from \( d_{k-1} \); (2) \textbf{Relevance Monotonicity:} The extent to which the retrieval model enforces a strict ordering such that \( S(q_k, d_j) > S(q_k, d_{j+1}) \) for all \( j \), ensuring that documents satisfying more conditions are ranked higher; and (3) \textbf{Format Invariance:} The stability of retrieval performance under transformations of query representation, quantified by discrepancies in ranking outcomes across query formats.

\subsection{Domain Selection}

To construct the \textsc{MultiConIR} dataset, we meticulously selected five domains—Books, Movies, People, Medical Cases, and Legal Documents—each chosen for its practical significance and inherent suitability for evaluating multi-condition retrieval capabilities. 
 
\textbf{Books \& Movies:} These domains represent common consumer searches where nuanced preferences are expressed by combining structured attributes (e.g., genre, creator, year, cast) with narrative elements (e.g., \textit{plot details, thematic content like “a story about time travel”}). Effective retrieval demands semantic understanding beyond simple keyword matching to process multifaceted queries, such as \textit{"an action film directed by Christopher Nolan, starring Leonardo DiCaprio, released after 2010, with an intense chase scene."}

\textbf{People:} Queries about individuals frequently rely on partial or vague information, such as notable achievements or specific traits. An example query could be \textit{“a Nobel laureate in Physics who studied black holes.”} These searches demand that IR systems effectively handle incomplete data and infer connections between various attributes to identify the correct individual.

\textbf{Medical Case \& Legal Document:} The medical case and legal document domains offer more practical and application-driven use cases. In the medical domain, doctors often rely on retrieval systems to reference historical cases to support diagnostic decisions. A typical query might include multiple conditions, such as \textit{“Find a case that meets the following conditions: 1) middle-aged female patient; 2) hospitalized for breathing difficulties; 3) has a history of antibiotic allergies; 4) has a family history of peanut allergies.”} Similarly, in the legal domain, retrieval users often seek case law with high similarity to ongoing cases, which requires matching various legal and factual attributes in historical court decisions. These complex queries require IR systems to perform fine-grained condition matching and understand the interdependencies between various factors.

\subsection{Dataset Construction Pipeline}

\begin{figure*}[!th]
  \includegraphics[width=\linewidth]{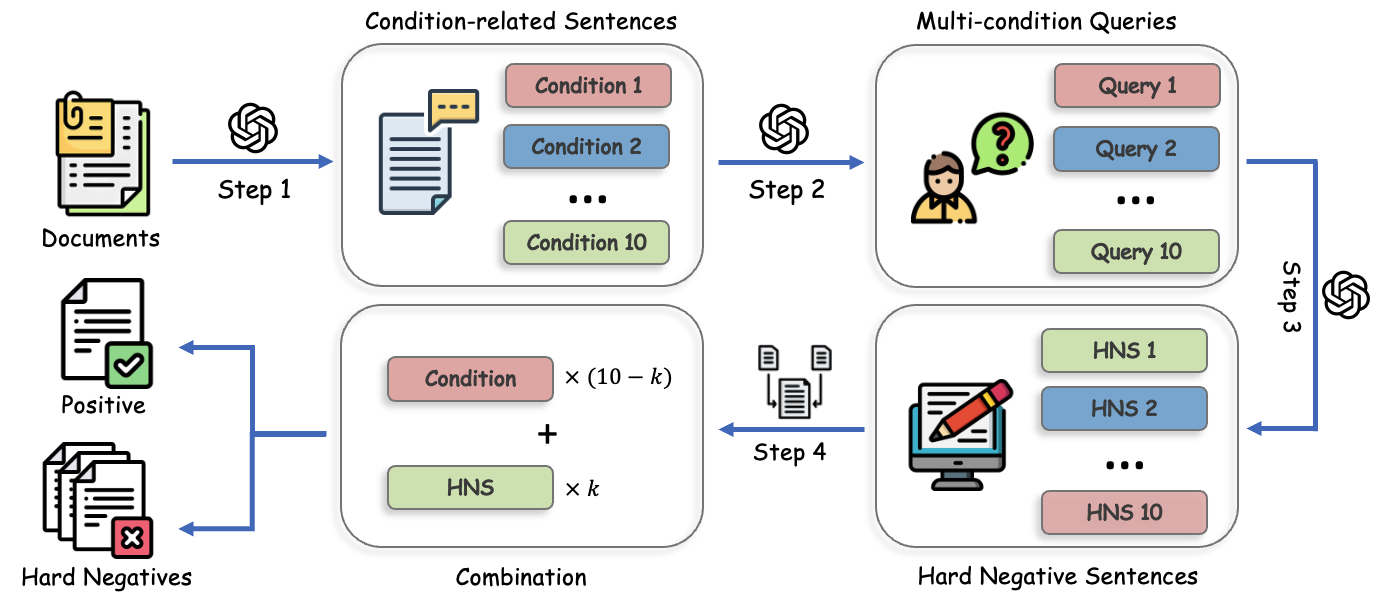} 
  \caption {\small \textsc{MultiConIR} Dataset Construction Pipeline: (1) relevant condition sentences are extracted from documents; (2) these conditions are then used to generate multi-condition queries; (3) hard negative (HN) versions of the condition sentences are created; and (4) positive documents and progressively challenging HN documents are assembled from these elements.}
  \label{fig:pipeline}
\end{figure*}

To construct \textsc{MultiConIR}, we design a multi-step data generation framework. As shown in Fig.~\ref{fig:pipeline}, this pipeline is highly adaptable across multiple domains, enabling the generation of queries and hard negative (HN) documents that progressively satisfy 1 to 10 conditions.
To preserve dataset integrity and mitigate the generalization issues associated with fully synthetic datasets~\citep{li2023synthetic, wang2024generateddatahelpcontrastive}, we employ LLM-based generation (GPT-4o) exclusively for modifying sentences within hard negatives, rather than altering entire documents.\footnote{Fully LLM-generated datasets may introducing inherent linguistic biases of the underlying LLMs, and lacks the contextual richness and complexity in real-world retrieval~\citep{shumailov2024ai}. To mitigate these issues, we restricted LLM interventions to modifying only condition sentences rather than entire document. We further discuss this problem in Appendix~\ref{Appendix:ai data}.}  
The data creation process consists of the following steps, with detailed prompt templates provided in Appendix~\ref{sec:prompt}:

\textbf{Step 1:} \textbf{Condition Sentences Extraction}.  
For each source document $d$, we issue a structured prompt to GPT-4o that (1) reads the entire text, (2) identifies ten non-overlapping conditions that describe key features expressed in $d$, and (3) return ten original sentences, where every sentence is semantically complete, and expresses a distinct condition. Documents yielding fewer than ten qualified sentences are discarded. The final ten sentences constitute the condition set $C(d)$, which we later recombine into queries of varying complexity and use as the ground-truth positive for multi-condition retrieval experiments.
% To capture the fine-grained constraints for multi-condition retrieval, we prompt GPT-4o to extract ten key sentences from each document, where each sentence represents a distinct and semantically complete condition. 

\textbf{Step 2:} \textbf{Query Generation}.  
Given the ten-sentence condition set $C(d)$, we prompt GPT-4o to synthesise a hierarchy of ten queries $\{ q_1,q_2,...,q_{10}\}$. Each $q_k$ incorporating the first $k$ conditions. To enhance linguistic diversity, For every $q_k$ we request two forms: (1) Instruction-style $q_k^{\text{inst}}$: a bullet-like template (\textit{“Find a document that satisfies: (1)… (2)…”)} for explicit parsing.  (2) Descriptive-style $q_k^{\text{des}}$: embedding conditions naturally within a coherent sentence.
% This dual-format design ensures **model robustness to query style variations**.  

\textbf{Step 3:} \textbf{Hard Negative Sentence Construction}.  
% To create hard negatives, we prompt GPT-4o to modify extracted condition sentences into Hard Negative Sentences (HNS) that subtly deviate from the original meaning while preserving lexical plausibility. Unlike naive negation, these modifications introduce semantic shifts that retain fluency but fail to satisfy the original retrieval conditions, thereby making retrieval harder.
% We experimented with two approaches for HNS construction: (1) modifies key information (applied to books, movies, medical cases, and legal documents); (2) retains keywords but adds dummy information (used for the people dataset). \footnote{Retrievers are more robust when dummy information is added; Modifying critical information is challenging for both retrievers and rerankers. The differences between these two approaches are further discussed in Appendix~\ref{Appendix:different HN}. }
For each condition sentence \(c_i \in C(d)\) we instruct \textbf{GPT-4o} to produce a semantically divergent yet fluently written variant \(h_i\) that no longer satisfies the original constraint.  The rewrite must preserve overall length and style while introducing either (i) a subtle alteration of a critical fact (applied to \emph{books}, \emph{movies}, \emph{medical cases}, and \emph{legal documents}) or (ii) an innocuous clause that injects misleading information without changing the existing keywords (used for the \emph{people} corpus). \footnote{Retrievers are more robust when adding misleading information; Modifying critical facts is challenging for both retrievers and rerankers. A detailed comparison of these two strategies is given in Appendix~\ref{Appendix:different HN}.}  The ten variants form the hard-negative set \(\mathrm{HNS}(d)=\{h_1,\dots,h_{10}\}\).

\textbf{Step 4:} \textbf{Hard Negative Document Generation}.  
Starting from the ten-sentence condition set \(C(d)=\{c_1,\dots ,c_{10}\}\) and its hard-negative counterparts \(\mathrm{HNS}(d)=\{h_1,\dots ,h_{10}\}\), we build an ordinal ladder of document variants.  
The \emph{positive} document \(d^{+}\) contains the full sequence \([c_1,\dots ,c_{10}]\).  
For each \(k\in\{1,\dots ,10\}\) we generate a HN document
$d_k^{-}= \bigl[h_1,\dots ,h_k,\;c_{k+1},\dots ,c_{10}\bigr],$
i.e., the first \(k\) conditions are replaced by their semantically perturbed versions while the remaining \(10-k\) conditions stay intact.  
This yields a controlled degradation chain 
\(d^{+}\!\rightarrow d_1^{-}\!\rightarrow\cdots\!\rightarrow d_{10}^{-}\), 
ranging from a single violated constraint to a completely adversarial variant.  
\emph{Coupled with the hierarchical queries \(\{Q_k\}_{k=1}^{10}\), the corpus enables fine-grained evaluation of retrieval performance under progressively stricter condition sets.}

\begin{figure}
    \centering
    \includegraphics[width=1\linewidth]{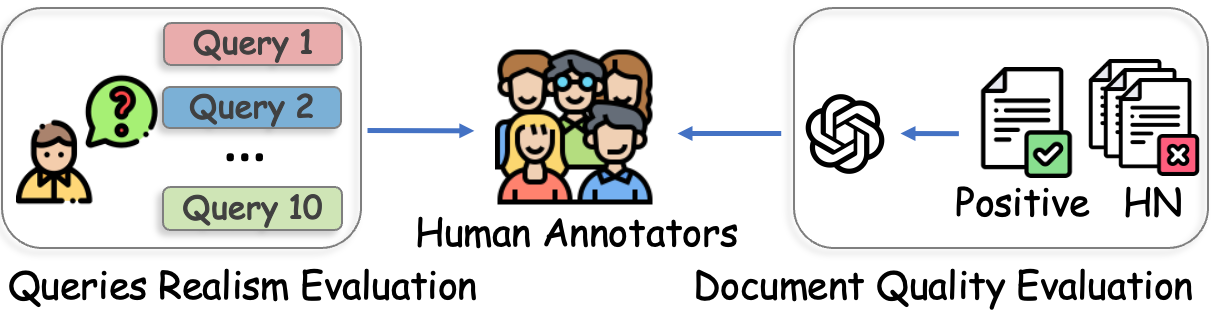}
    \caption{\small Benchmark quality evaluation framework of \textsc{MultiConIR}. Query realism was assessed by human annotators. Label accuracy involved initial GPT-4o filtering, followed by a final human double-check.
}
    \label{fig:quality}
\end{figure}

\paragraph{Benchmark Quality Assurance.}
The reliability of \textsc{MultiConIR} was audited on two complementary fronts (Fig.~\ref{fig:quality}).  
\textbf{(1) Query validity.}  
From each of the five domains we randomly sampled 100 multi-condition queries (500 in total) and asked ten trained annotators to judge whether each query was \emph{natural, precise,} and \emph{contextual plausibility (valid vs. unvalid)}.  The resulting inter-annotator agreement reached \(93.7\%\), with Fleiss’~\(\kappa = 0.84\).  \textbf{(2) Document–label validity.}  
To detect false positives/negatives, we first applied \textit{GPT-4o} to the \emph{entire} corpus: the model verified for every document variant \(d_k^{-}\) whether exactly \(k\) of the ten conditions were satisfied, and we discarded mismatched instances.  We then drew another 100 document–query pairs per domain (500 total) for manual spot-checking; two independent annotators reviewed each pair and a third adjudicated disagreements, yielding a residual error rate of \(2.4\%\).
After these filtering steps the final benchmark sizes for each domain are summarised in Table~\ref{tab:domain_sources}, and the full evaluation protocol is detailed in Appendix~\ref{sec:dataset quality}.

% \textbf{Benchmark Quality Assurance.} After constructing the MultiConIR dataset, we implemented a comprehensive evaluation framework to ensure the benchmark’s quality. As shown in Fig.\ref{fig:quality}, we employed multiple human annotators to assess the realism of multi-condition queries. Results demonstrate a high inter-annotator agreement (500 samples are randomly selected, 10 annotators achieve a IAA = 93.7\%). For document label quality, we initially utilized GPT-4o to filter and retain samples where the labels aligned with the model’s evaluations. Subsequently, human annotators conducted a double-check to confirm the accuracy of these labels. Detailed evaluation procedures are presented in Appendix \ref{sec:dataset quality}. In Table \ref{tab:domain_sources}, we present the benchmark statistics.

\begin{table*}[htbp]
  \centering
  \resizebox{=1\textwidth}{!} % Retaining original resizebox syntax
  {%
    \begin{tabular}{lcccccccc} % Adjusted to 7 columns: l for Domain, c for others
    \hline
    \multirow{2}{*}{\textbf{Domain}} & \multicolumn{3}{c}{\textbf{Number}} & \multicolumn{2}{c}{\textbf{Avg. Length}} & \multirow{2}{*}{\textbf{Source}} & \multirow{2}{*}{\textbf{License}} & \multirow{2}{*}{\textbf{Example}}\\
    \cline{2-4} \cline{5-6} % Partial lines for sub-headers
    & $D^+$ & $Q$ & $D^-$ & $Q$ & $D$ & \\
    \hline
    People & 420  & 4200 & 4200 & 31.2 & 225.2 & People Wikipedia Dat\citep{mahajan_people_wikipedia_data} & CC0 1.0 & Table \ref{tab:people_ex}\\
    Books & 482   & 4820 & 4820 & 25.6 & 235.3  & Books Dataset \citep{rustamov_books_dataset} & CC0 1.0 & Table \ref{tab:book_ex}\\
    Movies & 500  & 5000 & 5000 & 24.8 & 184.6  & Wikipedia Movie Plots \citep{robischon_wikipedia_movie_plots} & CC BY-SA 4.0 &Table \ref{tab:movie_ex}\\
    Medical Case & 479   & 4790 & 4790 & 28.4& 212.1  & Medical Transcription \citep{hpe2023medicalcases} & CC0 1.0 &Table \ref{tab:medical_ex}\\
    Legal Document & 426  & 4260 & 4260 & 34.1 & 302.2  & LexGLUE \citep{chalkidis-etal-2022-lexglue} & CC BY 4.0 &Table \ref{tab:legal_ex}\\
    \hline
    \end{tabular}%
    }
    \caption{\small Data statistics of MutiConIR. For each dataset, we show the number of positive documents ($D^+$
    ), queries ($Q$) and hard negative documents($D^-$), and the average length (in words) of queries and documents, and the source dataset of each domain.} 
    \label{tab:domain_sources}%
\end{table*}%

\subsection{Evaluation Metrics}
Conventional IR metrics—e.g., Precision@1, NDCG@\(k\)—only confirm whether a highly relevant document appears early, but they \emph{cannot} distinguish how well a model orders candidates that satisfy different numbers of query conditions.  
We therefore propose \textbf{Win Rate}: the proportion of pairwise comparisons in which a candidate that fulfils more conditions ranks above one that fulfils fewer. We further discuss the difference of Win Rate and traditional IR metrics in the Appendix \ref{sec:metric_discussion} 
\paragraph{Complexity Robustness}   

Queries range from \texttt{query1} to \texttt{query10}, each progressively incorporating 1 to 10 conditions. The candidate set comprises a \texttt{Positive} document that fully satisfies all conditions and a \texttt{HN1} document, which is derived from the positive by modifying a single condition.
Complexity robustness is measured using \textbf{Win Rate (WR)} \footnote{In Complexity Robustness evaluation, $\text{WR}_{k}$ and Precision@1 are numerically equivalent. }under various $k$, defined as:
\begin{equation*}
    \text{WR}_{k} = \frac{1}{N} \sum_{i=1}^{N} \mathbf{1} (S(q_k, d^+) > S(q_k, d_{k-1})),
\end{equation*}
where $S(q_i, d^+)$ and $S(q_i, d^-)$ denote similarity scores for the positive document and hard negative.

\paragraph{Relevance Monotonicity}
%This task investigates the model’s ability to distinguish subtle semantic differences in a multi-condition setting. 
The query is fixed as \texttt{query10} (containing all 10 conditions), while the candidate set includes one \texttt{positive} and ten hard negatives (\( d_0 -d_{9}\)), each containing 0–9 conditions.

We evaluate performance using $\text{WR}_{k, k-1}$
% \footnote{In Relevance Monotonicity evaluation, Precision@1 limited to top-1 accuracy, overlooks hierarchical relevance among partially relevant candidates;  $\text{WR}_{k, k-1}$ more effectively captures monotonic relevance as satisfied conditions increase.} 
between adjacent hard negatives:
\begin{equation*}
    \begin{split}
        \text{WR}_{k, k-1} &= \frac{1}{N} \sum_{i=1}^{N} \mathbf{1} \big( S(q_{10}, d_k) \\
        &\quad > S(q_{10}, d_{k-1}) \big),
    \end{split}
\end{equation*}

\paragraph{Format invariance.}
We compare two query formats:  
(1) Instruction-style, which explicitly lists conditions (e.g., \textit{Find a movie that meets the following conditions: 1. Action genre, 2. Directed by James Cameron}).  
(2) Descriptive-style, which integrates conditions into a natural query (e.g., \textit{Find an action movie directed by James Cameron}).  

To quantify ranking variability between query styles, we define the \textbf{Flip Rate (FR)}:
\begin{equation*}
    \begin{split}
        \text{FR} = \frac{1}{N} \sum_{i=1}^{N} \mathbf{1} \big( \mathbf{1}(S_{\text{inst}}(q_{10}, d_{k}) > S_{\text{inst}}(q_{10}, d_{k-1})) \\
        \neq \mathbf{1}(S_{\text{desc}}(q_{10}, d_{k}) > S_{\text{desc}}(q_{10}, d_{k-1})) \big),
    \end{split}
\end{equation*}
where \( S_{\text{inst}} \) and \( S_{\text{desc}} \) denote similarity scores under instruction-style and descriptive-style queries. The indicator function returns 1 if the ranking order of positive and hard negative documents changes between query styles and 0 otherwise. A higher FR indicates greater sensitivity to query formulation.

\section{Experiments}
We evaluate 15 representative retrieval and reranking models from diverse architectures and varying model sizes, including sparse retrieval model: BM25~\citep{robertson2009bm25}; BERT-based retrieval models: gte-large-en-v1.5~\citep{li2023towards} and jina-embeddings-v3 \citep{sturua2024jina}; LLM-based retrieval models: NV-Embed-v2~\citep{lee2024nv}, bge-en-icl~\citep{li2024makingtextembeddersfewshot}, gte-Qwen2-7B-instruct~\citep{li2023towards}, gte-Qwen2-1.5B-instruct~\citep{li2023towards}, e5-mistral-7b-instruct~\citep{wang-etal-2024-improving-text}, GritLM-7B~\citep{muennighoff2024generative}, LLM2Vec~\citep{behnamghader2024llm2veclargelanguagemodels}; Point-wise reranking models: bge-reranker-v2-m3~\citep{chen2024bge}, bge-reranker-v2-gemma~\citep{chen2024bge}, FollowIR-7B~\citep{weller2024followir}; Fine-tuned list-wise reranker: RankZephyr~\citep{pradeep2023rankzephyr}; Advanced LLM for zero-shot ranking: GPT-4o~\citep{OpenAI2024helloGPT4o}. Details of each model are provided in Appendix~\ref{sec:details of models}.

\begin{table*}[!t]
  \centering
  \resizebox{=\textwidth}{!}{ % Kept original resizebox syntax as per your input
    \begin{tabular}{cccccccccccc}
    \toprule
    Model & Query1 & Query2 & Query3 & Query4 & Query5 & Query6 & Query7 & Query8 & Query9 & Query10 & \textbf{Decline} \\
    \midrule
    \multicolumn{12}{c}{Sparse Retriever} \\
    \midrule
    \rowcolor[rgb]{ 1,  .953,  .949} BM25  & 28.59 & 34    & 33.14 & 36.53 & 37.38 & 37.95 & 38.08 & 38.41 & 38.86 & 39.87 & $\downarrow$-11.28 \\
    \midrule
    \multicolumn{12}{c}{ Dense Retriever} \\
    \midrule
    \rowcolor[rgb]{.949,  .949,  1} jina-embeddings-v3 & 76.09 & 71.26 & 71.84 & 71.04 & 65.59 & 65.65 & 64.75 & 64.24 & 64.62 & 60.71 & $\downarrow$15.38 \\
    
    \rowcolor[rgb]{ .949,  .949,  1} gte-large-en-v1.5 & 75.87 & 77.26 & 73.79 & 70.22 & 70.71 & 67.40  & 67.53 & 64.97 & 65.36 & 61.36 & $\downarrow$14.51 \\
    
    \rowcolor[rgb]{ .949,  .949,  1} NV-Embed-v2 & 80.53 & 80.32 & 78.81 & 75.70  & 75.68 & 72.61 & 73.28 & 71.54 & 70.00     & 68.02 & $\downarrow$12.51 \\
    
    \rowcolor[rgb]{ .949,  .949,  1} bge-en-icl & 83.42 & 80.65 & 78.44 & 76.77 & 74.54 & 73.00    & 74.23 & 73.25 & 69.70  & 68.00    & $\downarrow$15.42 \\
    
    \rowcolor[rgb]{ .949,  .949,  1} gte-Qwen2-7B-instruct & 70.75 & 72.22 & 69.99 & 68.51 & 65.20  & 63.53 & 62.22 & 62.20  & 59.15 & 56.17 & $\downarrow$14.58 \\
    
    \rowcolor[rgb]{ .949,  .949,  1} gte-Qwen2-1.5B-instruct & 73.64 & 74.97 & 72.23 & 71.37 & 69.94 & 67.46 & 66.92 & 64.64 & 63.65 & 58.68 & $\downarrow$14.96 \\
    
    \rowcolor[rgb]{ .949,  .949,  1} e5-mistral-7b-instruct & 75.05 & 70.85 & 68.18 & 67.45 & 63.70  & 61.60  & 59.70  & 59.07 & 57.85 & 58.12 & $\downarrow$16.93 \\
    
    \rowcolor[rgb]{ .949,  .949,  1} GritLM-7B & 82.08 & 80.32 & 78.38 & 76.40  & 76.40  & 73.50  & 75.69 & 74.62 & \underline{74.53} & \underline{75.95} & \textbf{$\downarrow$6.13} \\
    
    \rowcolor[rgb]{ .949,  .949,  1} LLM2Vec & 83.13 & 77.42 & 75.49 & 75.48 & 72.49 & 72.56 & 70.56 & 70.73 & 68.71 & 67.00    & $\downarrow$16.13 \\
    \midrule
    \multicolumn{12}{c}{Fine-tuned Reranker} \\
    \midrule
    \rowcolor[rgb]{  .949,  1,  .949} bge-reranker-v2-m3 & 87.14 & 85.56 & 78.62 & 76.05 & 74.29 & 68.41 & 67.86 & 59.48 & 55.59 & 44.87 & $\downarrow$42.27 \\
    
    \rowcolor[rgb]{ .949,  1,  .949} bge-reranker-v2-gemma & 91.07 & 90.02 & 86.70  & 84.99 & 83.17 & 79.00    & 75.89 & 72.29 & 67.11 & 56.09 & $\downarrow$34.98 \\
    
    \rowcolor[rgb]{  .949,  1,  .949} followIR & 83.41 & 79.72 & 76.25 & 74.60  & 70.12 & 67.94 & 62.62 & 55.93 & 48.59 & 43.52 & $\downarrow$39.89 \\

    \rowcolor[rgb]{  .949,  1,  .949} RankZephyr  & \underline{92.72} & \underline{90.29} & \underline{88.38} & \underline{87.69} & \underline{84.57} & \underline{80.88} & \underline{78.93} & \underline{75.99} & 72.39 & 66.84 & $\downarrow$25.88 \\
    \midrule 
    \multicolumn{12}{c}{Zero-shot LLM for Ranking} \\
    \midrule
    \rowcolor[rgb]{  .949,  .949,  .949} GPT-4o  & \textbf{95.49} & \textbf{94.89} & \textbf{93.71} & \textbf{92.11} & \textbf{90.81} & \textbf{89.08} & \textbf{88.43} & \textbf{88.08} & \textbf{86.82} & \textbf{85.26} & $\downarrow$\underline{9.23} \\
    \bottomrule
    \end{tabular}%
    }
      \caption{Impact of increasing condition quantity in queries on average Win Rate (Task 1).
The Decline reflects the degree of Win Rate reduction from query1 to query10.}
  \label{tab:task1}%
\end{table*}%

\subsection{Results for Complexity Robustness}

Table~\ref{tab:task1} presents the average Win Rate scores for evaluating complexity robustness across five datasets. 
The results reveal several notable trends:

\paragraph{Performance decline with more conditions}  
As the number of conditions in the query increases, the performance of both retrieval and reranking models declines. This suggests that with more conditions, models struggle to accurately distinguish between positives and HNs. Among all models, \emph{GPT-4o} maintained the highest win rate from Query1 to Query10, with its performance declining by 9.23\%. \emph{GritLM-7B} exhibits the lowest performance degradation of  6.13\%.The remaining models all exceeded 10\% decline.

\paragraph{Rerankers exhibit steeper performance drop}  
 As shown in Table~\ref{tab:task1}, fine-tuned rerankers outperform retrievers with single-condition queries. However, as the number of conditions increases, their performance declines more sharply. \emph{Eventually, rerankers even fell behind some retrievers.} The Win Rates for all rerankers declined by over 25\%, with an average decline of 35.76\%. For retriever models, the average decline was 14.06\%.
 
\subsection{Results for Relevance Monotonicity}
\begin{figure}
    \centering
    \includegraphics[width=1\linewidth]{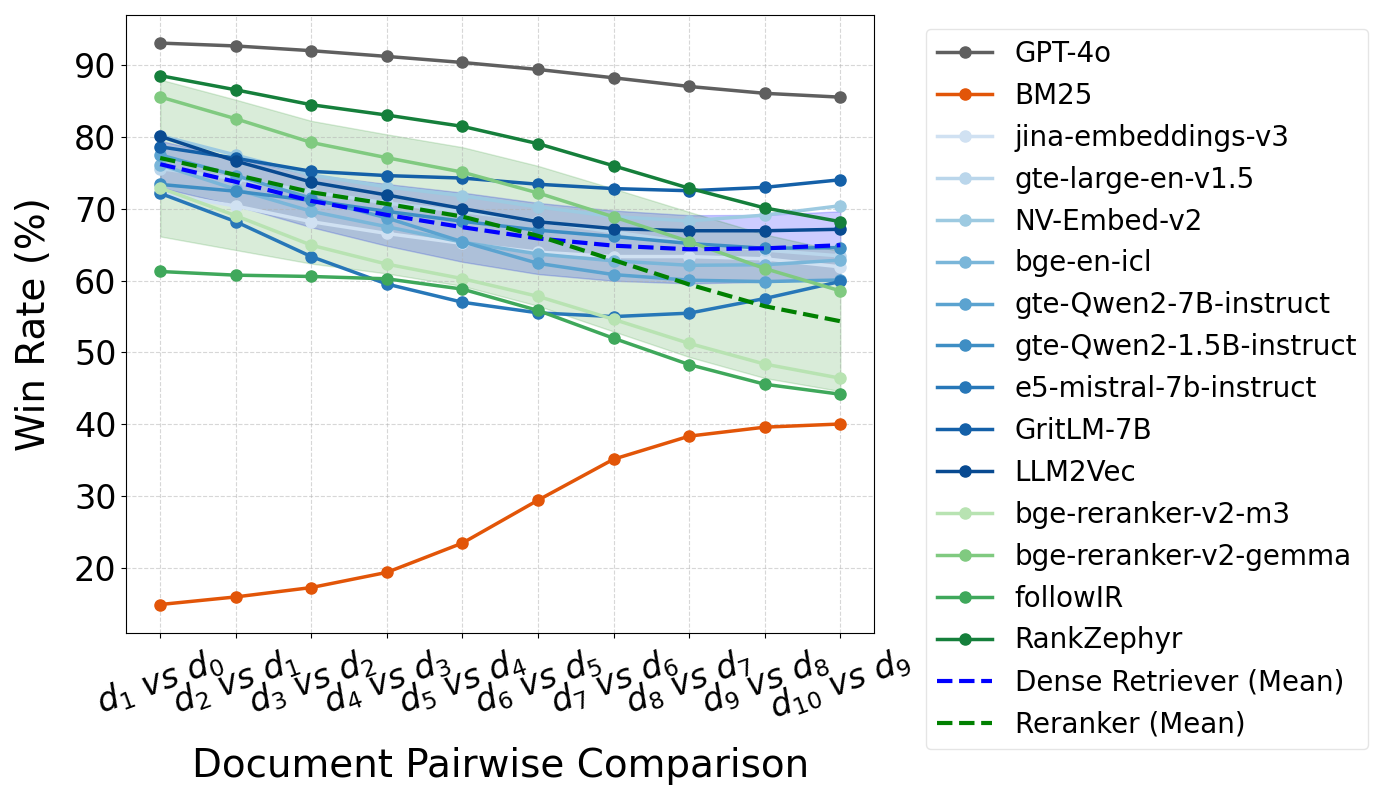}
    \caption{\small Relevance Monotonicity Distinction. Win Rate reflects the success rate between documents satisfying different numbers of conditions under a multi-condition query.
}
    \label{fig:task2-winrate}
\end{figure}

Fig.~\ref{fig:task2-winrate} illustrates the trend of average $\text{WR}_{k, k-1}$ in the multi-condition retrieval setting of Task 2, which evaluates the model’s ability to distinguish the relevance hierarchy among documents with varying conditions. The complete results are provided in Table \ref{tab:task2}. Several key observations can be made:

\paragraph{Relevance monotonicity struggle} As documents become increasingly hard (i.e., satisfying more conditions in the query), it becomes harder for retrieval and reranking models to accurately distinguish  \( d_k \) and \( d_{k-1} \), leading to a decline in 
Win Rate performance. This failure emphasizes the challenge of preserving relevance monotonicity in multi-condition retrieval settings and highlights a gap in current model capabilities when handling complex queries.

\paragraph{Sensitive to exact match and  complete mismatches} 
We observe a slight upward trend at the end of Win Rate curves for most dense retrievers, likely due to their contrastive learning-based training. Traditional contrastive learning treats retrieval as a binary task, pulling query-positive pairs closer while pushing negatives further apart, without accounting for partial matches. As a result, dense retrievers perform more reliably in clear-cut “exact match” or “complete mismatch” cases.

\subsection{Results for Format Invariance} 
Table~\ref{tab:flip rate} presents the Flip Rate induced by query format variations. \emph{GPT-4o showed the lowest flip rate of 6.98\%, showcasing the robustness of advanced LLMs against variations in query style.} Additionally, most models exceed 10\%, indicating a substantial impact of query formulation on retrieval performance. 
Table \ref{tab:flip rate} presents the complete results of Format Invariance.
\begin{table}[!ht]
  \centering
  \resizebox{=0.49\textwidth}{!}{
    \begin{tabular}{ccccccc}
        \toprule
    Model & People & Books & Movies & Medical  & Legal & Avg. \\
    \midrule
    \multicolumn{7}{c}{Sparse Retriever} \\
    \midrule
    \rowcolor[rgb]{ 1,  .953,  .949} BM25  & 14.86 & 17.88 & 16.84 & 19.25 & 12.14 & 16.19 \\
    \midrule
    \multicolumn{7}{c}{Dense Retriever} \\
    \midrule
    \rowcolor[rgb]{ .949,  .949,  1} jina-embeddings-v3 & 10.55 & 8.65  & 10.24 & 14.72 & 13.10  & 11.45 \\
    
    \rowcolor[rgb]{ .949,  .949,  1} gte-large-en-v1.5 & 11.74 & 8.84  & 12.96 & 15.70  & 15.16 & 12.88 \\
    
    \rowcolor[rgb]{ .949,  .949,  1} NV-Embed-v2 & 10.17 & 8.80   & 7.52  & 10.17 & 8.94  & 9.12 \\
    
    \rowcolor[rgb]{ .949,  .949,  1} bge-en-icl & 13.48 & 12.74 & 15.18 & 19.81 & 14.44 & 15.13 \\
    
    \rowcolor[rgb]{ .949,  .949,  1} gte-Qwen2-7B-instruct & 12.71 & 15.56 & 13.62 & 16.37 & 17.56 & 15.16 \\
    
    \rowcolor[rgb]{ .949,  .949,  1} gte-Qwen2-1.5B-instruct & 12.38 & 13.26 & 10.48 & 16.81 & 12.51 & 13.09 \\
    
    \rowcolor[rgb]{ .949,  .949,  1} e5-mistral-7b-instruct & 9.17  & 9.92  & 8.20   & 10.75 & 12.25 & 10.06 \\
    
    \rowcolor[rgb]{ .949,  .949,  1} GritLM-7B & 8.52  & 5.35  & 8.32  & 8.98  & 9.86  & \underline{8.21} \\
    
    \rowcolor[rgb]{ .949,  .949,  1} LLM2Vec & 12.81 & 7.93  & 9.56  & 8.12  & 10.49 & 9.78 \\
    \midrule
    \multicolumn{7}{c}{Fine-tuned Reranker} \\
    \midrule
    \rowcolor[rgb]{ .949,  1,  .949} bge-reranker-v2-m3 & 42.40  & 32.22 & 34.82 & 28.35 & 31.24 & 33.81 \\
    \rowcolor[rgb]{ .949,  1,  .949} bge-reranker-v2-gemma & 27.50  & 18.94 & 16.52 & 13.42 & 24.41 & 20.16 \\
    \rowcolor[rgb]{ .949,  1,  .949} followIR & 35.81 & 31.60  & 23.70  & 25.07 & 28.43 & 28.92 \\
    \rowcolor[rgb]{ .949,  1,  .949} RankZephyr & 20.21 & 16.34 & 14.86 & 11.53 & 25.31 & 17.65\\  
    \midrule 
    \multicolumn{7}{c}{Zero-shot LLM for Ranking} \\
    \midrule
    \rowcolor[rgb]{  .949,  .949,  .949} GPT-4o & 7.21 & 4.22 & 6.78 & 6.32 & 10.35 & \textbf{6.98}\\  
    \bottomrule
    \end{tabular}%
    }
    \caption{\small Flip Rate for query format shift (Task 3). The Flip Rate reflects the win rate reversal when switching the query format from instruction-style to descriptive-style.}
  \label{tab:flip rate}%
\end{table}%

Dense retrieval models show relatively lower sensitivity than rerankers, with Flip Rates between 8\% to 16\%.  GritLM-7B (8.21\%), NV-Embed-v2 (9.12\%), and LLM2Vec (9.78\%) exhibit less variation. In contrast, \emph{reranking models show significantly higher sensitivity to query format changes}, with Flip Rates exceeding 20\%. The highest Flip Rate observed is 33.81\% for bge-reranker-v2-m3.

\begin{figure*}[!ht]
    \centering
    \begin{subfigure}{0.49\textwidth}
        \centering
        \includegraphics[width=\linewidth]{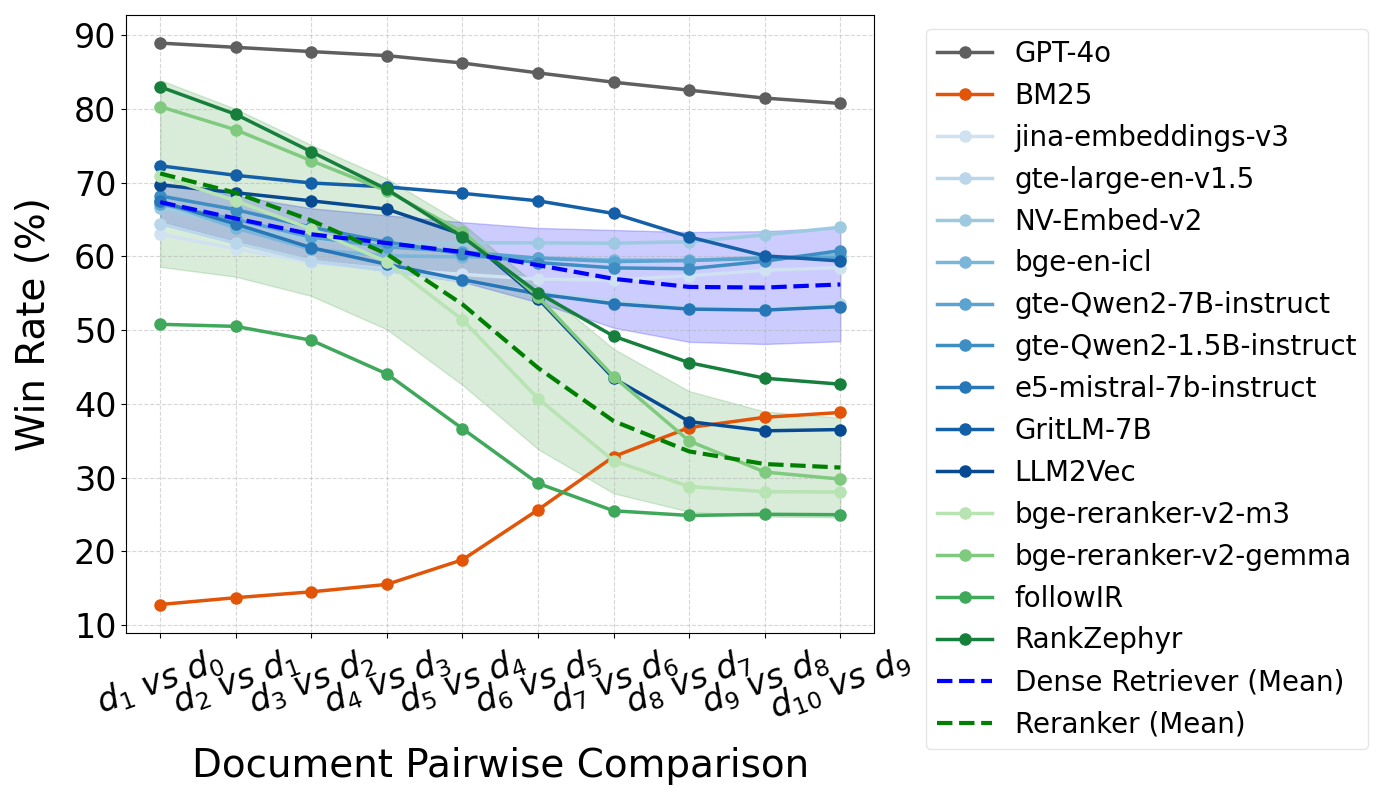}
        \caption{\small Retrieval Performance when padding to 512 words}
        \label{fig:task4-512}
    \end{subfigure}
    \hfill
    \begin{subfigure}{0.49\textwidth}
        \centering
        \includegraphics[width=\linewidth]{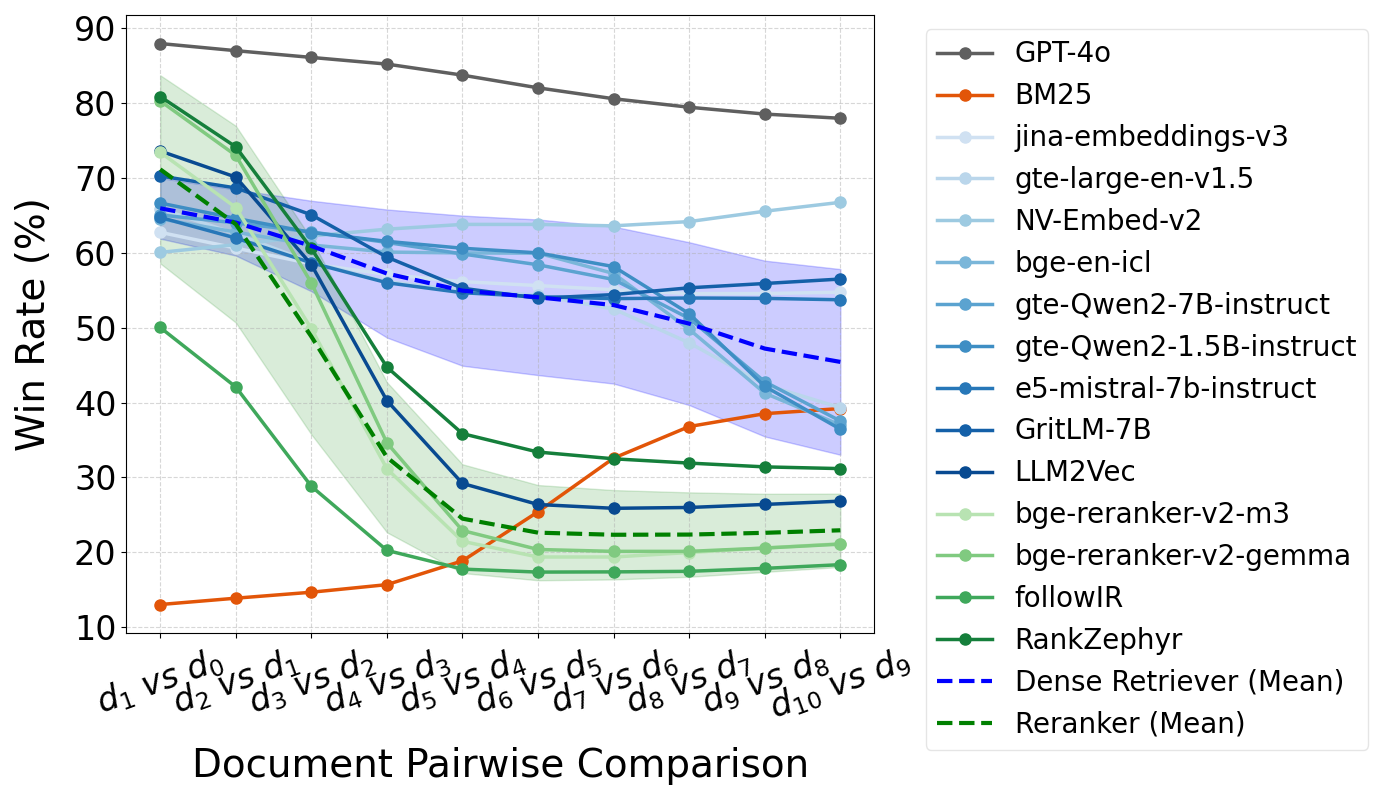}
        \caption{\small Retrieval Performance when padding to 1024 words}
        \label{fig:task4-1024}
    \end{subfigure}
    \caption{\small Retrieval performance when padding the document set to 512 words and 1024 words. Rerankers are highly sensitive to increases in document length, showing rapid performance degradation, whereas retrievers remain comparatively robust.}
    \label{fig:task4-length}
    % \vspace{-10pt}
\end{figure*}

\section{Analysis}

\subsection{Retrievers vs. Rankers}
Our experiments reveal notable differences between retrievers and rerankers across the three tasks. Fine-tuned rerankers exhibited excellent ranking performance under single-condition queries, but their efficacy rapidly diminished as the number of conditions increased. Retrievers demonstrate greater robustness under multi-condition queries and query style variations.

We hypothesize that one contributing factor to these performance disparities lies in the training datasets. Many dense retrieval models are trained on a mixture of retrieval-specific and general textual datasets~\citep{lee2024nv, behnamghader2024llm2veclargelanguagemodels, wang-etal-2024-improving-text}. Such diverse training enhances their generalization across various retrieval scenarios and query styles, which, in turn, improves their robustness against query complexity.

Beyond training data, we posit that the distinct input processing mechanisms also contribute to the observed performance differences. Retriever models typically employ a bi-encoder architecture, processing queries and documents independently. Conversely, rerankers, which process a concatenation of the query and document as a single input, appear more susceptible to input complexification—whether arising from an increase in query conditions or changes in query style.
To validate this speculation, we re-evaluated the Win Rate for the relevance monotonicity task under multi-condition queries using documents that padded to 512 and 1024 words. Results as shown in Fig.\ref{fig:task4-length} , revealed that fine-tuned rerankers are highly sensitive to such increases in document length, which further illustrates the sensitivity of rerankers to complex input. In contrast, retriever models demonstrated greater robustness to length modifications.

\subsection{Condition Position Impact on Focus}
\begin{figure*}[!ht]
    \centering
    \begin{subfigure}{0.32\textwidth}
        \centering
        \includegraphics[width=\linewidth]{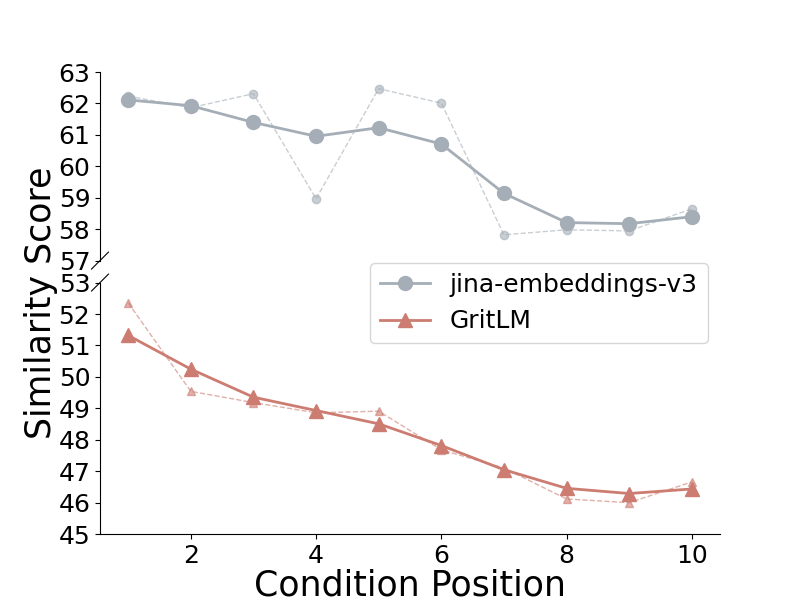} 
        \caption{Mean pooling}
        \label{fig:subfig2}
    \end{subfigure}
    \begin{subfigure}{0.32\textwidth}
        \centering
        \includegraphics[width=\linewidth]{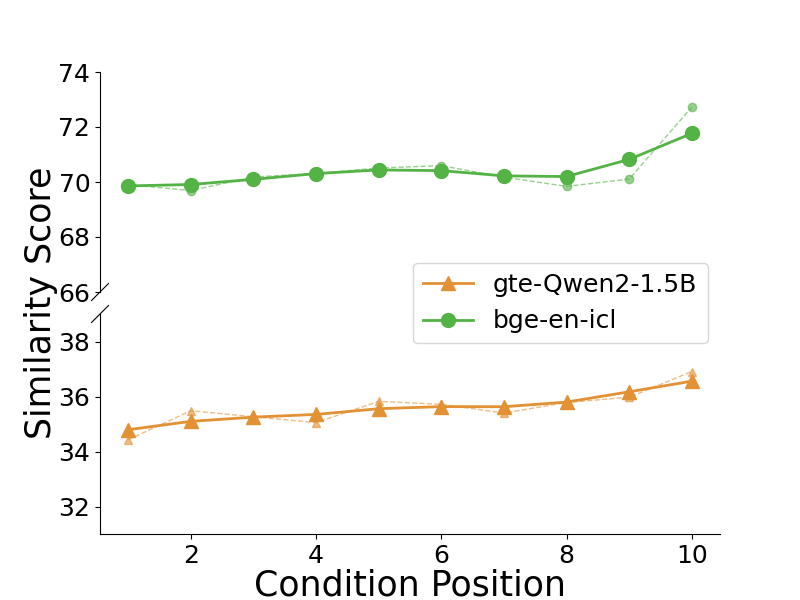} 
        \caption{\texttt{<EOS>} pooling}
        \label{fig:subfig3}
    \end{subfigure}
    \begin{subfigure}{0.32\textwidth}
        \centering
        \includegraphics[width=\linewidth]{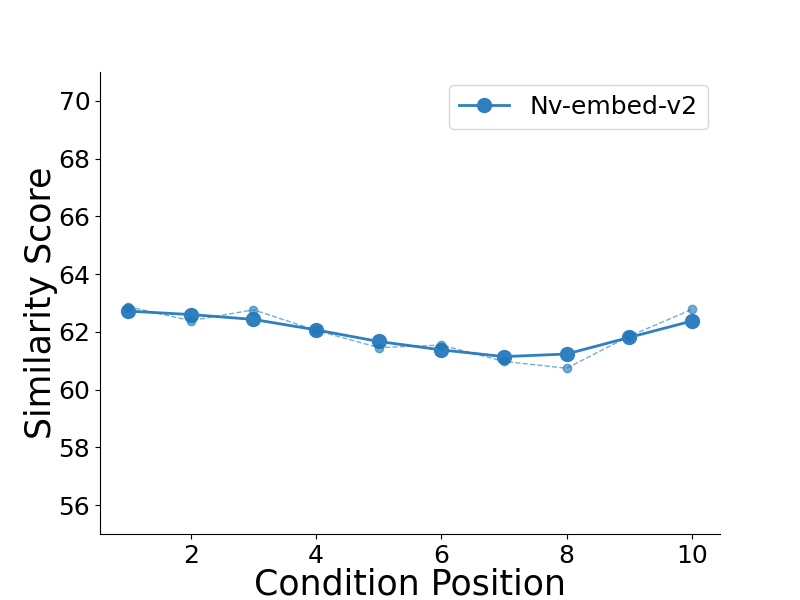} 
        \caption{Latent attention layer pooling}
        \label{fig:subfig1}
    \end{subfigure}
    \caption{\small Impact of condition position on different pooling methods. The condition is placed at different positions (1–10) in the query, with other positions filled by ten “[unused0]” tokens (example from Table~\ref{tab:book_ex}). The dashed line represents the original data, while the solid line shows the Gaussian-smoothed trend (kernel size = 1) for clarity.}
    \label{fig:pooling method}
    % \vspace{-10pt}
\end{figure*}

Our experimental findings indicate that the position of a condition significantly influences the model's subsequent similarity judgment. This phenomenon was observed consistently across both retriever and reranker models.

To illustrate position effects in retriever models, we conducted a targeted study on how relocating a single condition within the query influences its similarity score. We selected representative retriever models employing distinct pooling strategies: mean pooling, \texttt{<EOS>} pooling, and latent layer pooling. The results, as depicted in Fig.\ref{fig:pooling method}, revealed that models utilizing mean pooling tend to weight the early tokens most heavily: similarity drops steadily as the condition is shifted toward the tail of the query. \texttt{<EOS>} shows the opposite bias, emphasising the final tokens. Latent layer pooling heightened focus on both the beginning and end of the query, with comparatively less focus on the middle.

% Figure \ref{fig:cross-heatmap} shows the attention-score heat map produced by the cross-encoder reranker (bge-reranker-m3) for the input query–document pair. As we progressively add conditions to the query and compute the attention distribution between each condition and its corresponding segment in the document, we observe that the cross-encoder allocates attention unevenly across positions.
\begin{figure*}
    \centering
    \includegraphics[width=1\linewidth]{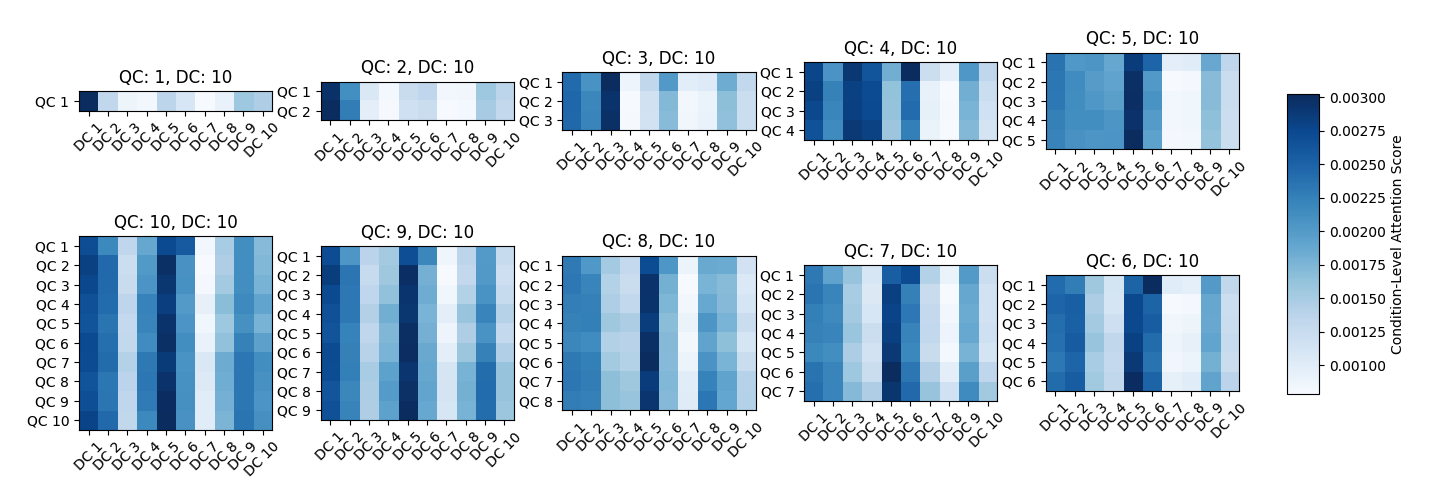}
    \caption{\small Attention heatmap of cross-encoder model (bge-reranker-m3).}
    \label{fig:cross-heatmap}
\end{figure*}

Similarly, for reranker models, we selected a cross-encoder model (bge-reranker-m3) to visualize the attention heatmap. Fig.\ref{fig:cross-heatmap}  shows a non-uniform distribution of attention across different token positions. This implies that these rerankers tend to assign differential focus to specific conditions or tokens within the concatenated query-document input, rather than distributing their attention uniformly across all elements.

\section{Conclusion}

In this work, we introduced \textsc{MultiConIR}, a novel benchmark designed to rigorously evaluate information retrieval models in realistic multi-condition scenarios, a critical area where existing evaluation frameworks are lacking. Through three specifically designed tasks—complexity robustness, relevance monotonicity, and query format sensitivity—conducted across five diverse domains. Experiments revealed that existing models struggle with multi-condition retrieval, with their performance degrading as the number of conditions increases; rerankers excel for single-condition queries but fail in multi-condition scenarios. Notably, rerankers are more sensitive to complex inputs.  GPT-4o outperforms specialised IR systems, exposing a performance gap in handling complex information needs.

Our findings highlight an urgent need for new modeling approaches and training paradigms specifically tailored for robust multi-condition understanding. \textsc{MultiConIR} serves as a valuable resource to drive this research, benchmark progress, and ultimately propel information retrieval systems towards a more sophisticated, human-like comprehension of complex information needs.

\section*{Limitations}
While \textsc{MultiConIR} provides a novel benchmark for evaluating retrieval models in multi-condition scenarios, several limitations should be acknowledged. First, our dataset relies on automated query generation and hard negative creation, which may introduce biases in condition representation despite efforts to ensure accuracy. These biases could affect retrieval models’ ability to distinguish fine-grained differences. Second, our evaluation focuses on retrieval tasks and does not cover reasoning-based retrieval or interactive search scenarios. Real-world systems often incorporate reranking, user feedback, and hybrid retrieval, which are not explicitly modeled. Lastly, our dataset does not fully consider query reformulation strategies or multi-turn retrieval, limiting its applicability to dynamic search environments. These limitations highlight the need for further research into multi-condition retrieval, particularly in addressing dataset biases, expanding evaluation scopes, and integrating retrieval with realistic user interactions.

\section*{Ethics Statement}

This study adheres to ethical standards in AI research, ensuring transparency and reproducibility in dataset construction and model evaluation while exclusively using publicly available pre-trained models for experiments.
Dataset Considerations: \textsc{MultiConIR} is built from publicly available sources and does not contain sensitive or personally identifiable information. Given its inclusion of medical and legal documents, we apply strict data filtering and safety measures to respect model safety constraints and prevent the generation of harmful or misleading content. Additionally, we recognize that automatically generated queries and hard negatives may introduce biases. Therefore, during dataset construction, we take measures to minimize the impact of inherent language model biases on retrieval tasks.
MultiConIR aims to advance multi-condition retrieval research while ensuring data fairness and ethical compliance. We encourage future research to further explore bias detection strategies in retrieval dataset, enhancing model fairness and reliability in diverse corpus environments.

\bibliography{custom}

\begin{thebibliography}{58}
\providecommand{\natexlab}[1]{#1}

\bibitem[{Asai et~al.(2023)Asai, Schick, Lewis, Chen, Izacard, Riedel, Hajishirzi, and Yih}]{asai2023task}
Akari Asai, Timo Schick, Patrick Lewis, Xilun Chen, Gautier Izacard, Sebastian Riedel, Hannaneh Hajishirzi, and Wen-tau Yih. 2023.
\newblock Task-aware retrieval with instructions.
\newblock \emph{ACL Findings}.

\bibitem[{BehnamGhader et~al.(2024)BehnamGhader, Adlakha, Mosbach, Bahdanau, Chapados, and Reddy}]{behnamghader2024llm2veclargelanguagemodels}
Parishad BehnamGhader, Vaibhav Adlakha, Marius Mosbach, Dzmitry Bahdanau, Nicolas Chapados, and Siva Reddy. 2024.
\newblock \href {https://arxiv.org/abs/2404.05961} {Llm2vec: Large language models are secretly powerful text encoders}.
\newblock \emph{Preprint}, arXiv:2404.05961.

\bibitem[{Berger et~al.(2000)Berger, Caruana, Cohn, Freitag, and Mittal}]{berger2000bridging}
Adam Berger, Rich Caruana, David Cohn, Dayne Freitag, and Vibhu Mittal. 2000.
\newblock Bridging the lexical chasm: statistical approaches to answer-finding.
\newblock In \emph{Proceedings of the 23rd annual international ACM SIGIR conference on Research and development in information retrieval}, pages 192--199.

\bibitem[{Carpineto and Romano(2012)}]{carpineto2012survey}
Claudio Carpineto and Giovanni Romano. 2012.
\newblock A survey of automatic query expansion in information retrieval.
\newblock \emph{Acm Computing Surveys (CSUR)}, 44(1):1--50.

\bibitem[{Chalkidis et~al.(2022)Chalkidis, Fergadiotis, Malakasiotis, Aletras, and Androutsopoulos}]{chalkidis-etal-2022-lexglue}
Ilias Chalkidis, Manos Fergadiotis, Prodromos Malakasiotis, Nikolaos Aletras, and Ion Androutsopoulos. 2022.
\newblock \href {https://doi.org/10.18653/v1/2022.acl-long.297} {{LexGLUE}: A benchmark dataset for legal language understanding in english}.
\newblock In \emph{Proceedings of the 60th Annual Meeting of the Association for Computational Linguistics (Volume 1: Long Papers)}, pages 4310--4330. Association for Computational Linguistics.

\bibitem[{Chen et~al.(2024)Chen, Xiao, Zhang, Luo, Lian, and Liu}]{chen2024bge}
Jianlv Chen, Shitao Xiao, Peitian Zhang, Kun Luo, Defu Lian, and Zheng Liu. 2024.
\newblock \href {https://arxiv.org/abs/2402.03216} {Bge m3-embedding: Multi-lingual, multi-functionality, multi-granularity text embeddings through self-knowledge distillation}.
\newblock \emph{Preprint}, arXiv:2402.03216.

\bibitem[{Devlin et~al.(2019)Devlin, Chang, Lee, and Toutanova}]{devlin2019bert}
Jacob Devlin, Ming-Wei Chang, Kenton Lee, and Kristina Toutanova. 2019.
\newblock \href {https://doi.org/10.18653/v1/N19-1423} {{BERT: Pre-training of Deep Bidirectional Transformers for Language Understanding}}.
\newblock In \emph{Proceedings of the 2019 Conference of the North American Chapter of the Association for Computational Linguistics: Human Language Technologies, Volume 1 (Long and Short Papers)}, pages 4171--4186. Association for Computational Linguistics.

\bibitem[{Ferraz et~al.(2024)Ferraz, Mehta, Lin, Chang, Oraby, Liu, Subramanian, Chung, Bansal, and Peng}]{ferraz2024llm}
Thomas~Palmeira Ferraz, Kartik Mehta, Yu-Hsiang Lin, Haw-Shiuan Chang, Shereen Oraby, Sijia Liu, Vivek Subramanian, Tagyoung Chung, Mohit Bansal, and Nanyun Peng. 2024.
\newblock Llm self-correction with decrim: Decompose, critique, and refine for enhanced following of instructions with multiple constraints.
\newblock \emph{arXiv preprint arXiv:2410.06458}.

\bibitem[{Han et~al.(2025)Han, Liu, and Huang}]{han2025attributestextualgenesleveraging}
Guangzeng Han, Weisi Liu, and Xiaolei Huang. 2025.
\newblock \href {https://arxiv.org/abs/2509.02040} {Attributes as textual genes: Leveraging llms as genetic algorithm simulators for conditional synthetic data generation}.
\newblock \emph{Preprint}, arXiv:2509.02040.

\bibitem[{He et~al.(2024)He, Zeng, He, Liang, and Xiao}]{he2024complex}
Qianyu He, Jie Zeng, Qianxi He, Jiaqing Liang, and Yanghua Xiao. 2024.
\newblock From complex to simple: Enhancing multi-constraint complex instruction following ability of large language models.
\newblock \emph{arXiv preprint arXiv:2404.15846}.

\bibitem[{Karpukhin et~al.(2020)Karpukhin, Oguz, Min, Lewis, Wu, Edunov, Chen, and Yih}]{karpukhin-etal-2020-dense}
Vladimir Karpukhin, Barlas Oguz, Sewon Min, Patrick Lewis, Ledell Wu, Sergey Edunov, Danqi Chen, and Wen-tau Yih. 2020.
\newblock \href {https://doi.org/10.18653/v1/2020.emnlp-main.550} {Dense passage retrieval for open-domain question answering}.
\newblock In \emph{Proceedings of the 2020 Conference on Empirical Methods in Natural Language Processing (EMNLP)}, pages 6769--6781, Online. Association for Computational Linguistics.

\bibitem[{Kwiatkowski et~al.(2019)Kwiatkowski, Palomaki, Redfield, Collins, Parikh, Alberti, Epstein, Polosukhin, Devlin, Lee et~al.}]{kwiatkowski2019natural}
Tom Kwiatkowski, Jennimaria Palomaki, Olivia Redfield, Michael Collins, Ankur Parikh, Chris Alberti, Danielle Epstein, Illia Polosukhin, Jacob Devlin, Kenton Lee, et~al. 2019.
\newblock Natural questions: a benchmark for question answering research.
\newblock \emph{Transactions of the Association for Computational Linguistics}, 7:453--466.

\bibitem[{Lee et~al.(2024)Lee, Roy, Xu, Raiman, Shoeybi, Catanzaro, and Ping}]{lee2024nv}
Chankyu Lee, Rajarshi Roy, Mengyao Xu, Jonathan Raiman, Mohammad Shoeybi, Bryan Catanzaro, and Wei Ping. 2024.
\newblock Nv-embed: Improved techniques for training llms as generalist embedding models.
\newblock \emph{arXiv preprint arXiv:2405.17428}.

\bibitem[{Li et~al.(2023{\natexlab{a}})Li, Liu, Xiao, and Shao}]{li2023makinglargelanguagemodels}
Chaofan Li, Zheng Liu, Shitao Xiao, and Yingxia Shao. 2023{\natexlab{a}}.
\newblock \href {https://arxiv.org/abs/2312.15503} {Making large language models a better foundation for dense retrieval}.
\newblock \emph{Preprint}, arXiv:2312.15503.

\bibitem[{Li et~al.(2024)Li, Qin, Xiao, Chen, Luo, Shao, Lian, and Liu}]{li2024makingtextembeddersfewshot}
Chaofan Li, MingHao Qin, Shitao Xiao, Jianlyu Chen, Kun Luo, Yingxia Shao, Defu Lian, and Zheng Liu. 2024.
\newblock \href {https://arxiv.org/abs/2409.15700} {Making text embedders few-shot learners}.
\newblock \emph{Preprint}, arXiv:2409.15700.

\bibitem[{Li et~al.(2023{\natexlab{b}})Li, Zhang, Zhang, Long, Xie, and Zhang}]{li2023towards}
Zehan Li, Xin Zhang, Yanzhao Zhang, Dingkun Long, Pengjun Xie, and Meishan Zhang. 2023{\natexlab{b}}.
\newblock Towards general text embeddings with multi-stage contrastive learning.
\newblock \emph{arXiv preprint arXiv:2308.03281}.

\bibitem[{Li et~al.(2023{\natexlab{c}})Li, Zhu, Lu, and Yin}]{li2023synthetic}
Zhuoyan Li, Hangxiao Zhu, Zhuoran Lu, and Ming Yin. 2023{\natexlab{c}}.
\newblock Synthetic data generation with large language models for text classification: Potential and limitations.
\newblock \emph{arXiv preprint arXiv:2310.07849}.

\bibitem[{Liu et~al.(2024)Liu, Chen, Ji, Zhou, Chen, and Wang}]{liu2024rag}
Wanlong Liu, Junying Chen, Ke~Ji, Li~Zhou, Wenyu Chen, and Benyou Wang. 2024.
\newblock Rag-instruct: Boosting llms with diverse retrieval-augmented instructions.
\newblock \emph{arXiv preprint arXiv:2501.00353}.

\bibitem[{Liu et~al.(2025)Liu, Xu, Yu, Lin, Ji, Chen, Xu, Wang, Shang, and Wang}]{liu2025qfft}
Wanlong Liu, Junxiao Xu, Fei Yu, Yukang Lin, Ke~Ji, Wenyu Chen, Yan Xu, Yasheng Wang, Lifeng Shang, and Benyou Wang. 2025.
\newblock Qfft, question-free fine-tuning for adaptive reasoning.
\newblock \emph{arXiv preprint arXiv:2506.12860}.

\bibitem[{Luan et~al.(2021)Luan, Eisenstein, Toutanova, and Collins}]{luan2021sparse}
Yi~Luan, Jacob Eisenstein, Kristina Toutanova, and Michael Collins. 2021.
\newblock \href {https://doi.org/10.1162/tacl\_a\_00369} {Sparse, dense, and attentional representations for text retrieval}.
\newblock \emph{Transactions of the Association for Computational Linguistics}, 9:329--345.

\bibitem[{Ma et~al.(2023)Ma, Wang, Yang, Wei, and Lin}]{ma2023finetuning}
Xueguang Ma, Liang Wang, Nan Yang, Furu Wei, and Jimmy Lin. 2023.
\newblock \href {https://arxiv.org/abs/2310.08319} {Fine-tuning llama for multi-stage text retrieval}.
\newblock \emph{Preprint}, arXiv:2310.08319.

\bibitem[{Mahajan(2017)}]{mahajan_people_wikipedia_data}
Sameer~S. Mahajan. 2017.
\newblock \href {https://www.kaggle.com/datasets/sameersmahajan/people-wikipedia-data} {People wikipedia data}.

\bibitem[{Mai et~al.(2024)Mai, Gauch, and Adams}]{mai2024setbert}
Quan Mai, Susan Gauch, and Douglas Adams. 2024.
\newblock Setbert: Enhancing retrieval performance for boolean logic and set operation queries.
\newblock \emph{arXiv preprint arXiv:2406.17282}.

\bibitem[{{MTSamples}(2025)}]{hpe2023medicalcases}
{MTSamples}. 2025.
\newblock \href {https://mtsamples.com/} {Medical cases classification tutorial dataset}.

\bibitem[{Muennighoff et~al.(2024)Muennighoff, Su, Wang, Yang, Wei, Yu, Singh, and Kiela}]{muennighoff2024generative}
Niklas Muennighoff, Hongjin Su, Liang Wang, Nan Yang, Furu Wei, Tao Yu, Amanpreet Singh, and Douwe Kiela. 2024.
\newblock \href {https://arxiv.org/abs/2402.09906} {Generative representational instruction tuning}.
\newblock \emph{Preprint}, arXiv:2402.09906.

\bibitem[{Muennighoff et~al.(2022)Muennighoff, Tazi, Magne, and Reimers}]{muennighoff2022mteb}
Niklas Muennighoff, Nouamane Tazi, Lo{\"\i}c Magne, and Nils Reimers. 2022.
\newblock Mteb: Massive text embedding benchmark.
\newblock \emph{arXiv preprint arXiv:2210.07316}.

\bibitem[{Nguyen et~al.(2016)Nguyen, Rosenberg, Song, Gao, Tiwary, Majumder, and Deng}]{nguyen2016ms}
Tri Nguyen, Mir Rosenberg, Xia Song, Jianfeng Gao, Saurabh Tiwary, Rangan Majumder, and Li~Deng. 2016.
\newblock Ms marco: A human-generated machine reading comprehension dataset.

\bibitem[{Nian et~al.(2024)Nian, Peng, Wang, and Fang}]{Nian2024WRAG}
Jinming Nian, Zhiyuan Peng, Qifan Wang, and Yi~Fang. 2024.
\newblock \href {https://arxiv.org/abs/2408.08444} {{W-RAG: Weakly Supervised Dense Retrieval in RAG for Open-domain Question Answering}}.
\newblock \emph{arXiv preprint arXiv:2408.08444}.

\bibitem[{Oh et~al.(2024)Oh, Lee, Ye, Shin, Jang, Jun, and Seo}]{oh2024instructir}
Hanseok Oh, Hyunji Lee, Seonghyeon Ye, Haebin Shin, Hansol Jang, Changwook Jun, and Minjoon Seo. 2024.
\newblock Instructir: A benchmark for instruction following of information retrieval models.
\newblock \emph{arXiv preprint arXiv:2402.14334}.

\bibitem[{{OpenAI}(2024)}]{OpenAI2024helloGPT4o}
{OpenAI}. 2024.
\newblock Hello gpt-4o.
\newblock \url{https://openai.com/index/hello-gpt-4o/}.
\newblock Accessed: 2025-05-19.

\bibitem[{Ponte and Croft(2017)}]{ponte2017language}
Jay~M Ponte and W~Bruce Croft. 2017.
\newblock A language modeling approach to information retrieval.
\newblock In \emph{ACM SIGIR Forum}, 2, pages 202--208. ACM New York, NY, USA.

\bibitem[{Pradeep et~al.(2023)Pradeep, Sharifymoghaddam, and Lin}]{pradeep2023rankzephyr}
Ronak Pradeep, Sahel Sharifymoghaddam, and Jimmy Lin. 2023.
\newblock \href {https://arxiv.org/abs/2312.02724} {Rankzephyr: Effective and robust zero-shot listwise reranking is a breeze!}
\newblock \emph{Preprint}, arXiv:2312.02724.

\bibitem[{Qin et~al.(2024)Qin, Song, Hu, Yao, Cho, Wang, Wu, Liu, Liu, and Yu}]{qin2024infobench}
Yiwei Qin, Kaiqiang Song, Yebowen Hu, Wenlin Yao, Sangwoo Cho, Xiaoyang Wang, Xuansheng Wu, Fei Liu, Pengfei Liu, and Dong Yu. 2024.
\newblock Infobench: Evaluating instruction following ability in large language models.
\newblock \emph{arXiv preprint arXiv:2401.03601}.

\bibitem[{Raffel et~al.(2020)Raffel, Shazeer, Roberts, Lee, Narang, Matena, Zhou, Li, and Liu}]{raffel2020exploring}
Colin Raffel, Noam Shazeer, Adam Roberts, Katherine Lee, Sharan Narang, Michael Matena, Yanqi Zhou, Wei Li, and Peter~J. Liu. 2020.
\newblock \href {http://jmlr.org/papers/v21/20-074.html} {Exploring the limits of transfer learning with a unified text-to-text transformer}.
\newblock \emph{Journal of Machine Learning Research}, 21(140):1--67.

\bibitem[{Ramos et~al.(2003)}]{ramos2003using}
Juan Ramos et~al. 2003.
\newblock Using tf-idf to determine word relevance in document queries.
\newblock In \emph{Proceedings of the first instructional conference on machine learning}, volume 242, pages 29--48. Citeseer.

\bibitem[{Robertson and Zaragoza(2009)}]{robertson2009bm25}
Stephen Robertson and Hugo Zaragoza. 2009.
\newblock \href {https://doi.org/10.1561/1500000019} {The probabilistic relevance framework: Bm25 and beyond}.
\newblock \emph{Foundations and Trends in Information Retrieval}, 3(4):333--389.

\bibitem[{Robischon(2018)}]{robischon_wikipedia_movie_plots}
Jonathan Robischon. 2018.
\newblock \href {https://www.kaggle.com/datasets/jrobischon/wikipedia-movie-plots} {Wikipedia movie plots}.

\bibitem[{Rustamov(2021)}]{rustamov_books_dataset}
Elvin Rustamov. 2021.
\newblock \href {https://www.kaggle.com/datasets/elvinrustam/books-dataset} {Books dataset}.

\bibitem[{Shi et~al.(2025)Shi, Xu, Zeqi, Zi, Wu, and Xu}]{shi-etal-2025-personax}
Yunxiao Shi, Wujiang Xu, Zhang Zeqi, Xing Zi, Qiang Wu, and Min Xu. 2025.
\newblock \href {https://doi.org/10.18653/v1/2025.findings-acl.300} {{P}ersona{X}: A recommendation agent-oriented user modeling framework for long behavior sequence}.
\newblock In \emph{Findings of the Association for Computational Linguistics: ACL 2025}, pages 5764--5787, Vienna, Austria. Association for Computational Linguistics.

\bibitem[{Shi et~al.(2024)Shi, Zi, Shi, Zhang, Wu, and Xu}]{Shi2024}
Yunxiao Shi, Xing Zi, Zijing Shi, Haimin Zhang, Qiang Wu, and Min Xu. 2024.
\newblock \href {https://doi.org/10.3233/FAIA240748} {Enhancing retrieval and managing retrieval: A four-module synergy for improved quality and efficiency in rag systems}.
\newblock In \emph{ECAI 2024}, pages 2258--2265. IOS Press.

\bibitem[{Shumailov et~al.(2024)Shumailov, Shumaylov, Zhao, Papernot, Anderson, and Gal}]{shumailov2024ai}
Ilia Shumailov, Zakhar Shumaylov, Yiren Zhao, Nicolas Papernot, Ross Anderson, and Yarin Gal. 2024.
\newblock Ai models collapse when trained on recursively generated data.
\newblock \emph{Nature}, 631(8022):755--759.

\bibitem[{Sturua et~al.(2024)Sturua, Mohr, Akram, G{\"u}nther, Wang, Krimmel, Wang, Mastrapas, Koukounas, Wang et~al.}]{sturua2024jina}
Saba Sturua, Isabelle Mohr, Mohammad~Kalim Akram, Michael G{\"u}nther, Bo~Wang, Markus Krimmel, Feng Wang, Georgios Mastrapas, Andreas Koukounas, Nan Wang, et~al. 2024.
\newblock jina-embeddings-v3: Multilingual embeddings with task lora.
\newblock \emph{arXiv preprint arXiv:2409.10173}.

\bibitem[{Su et~al.(2024)Su, Yen, Xia, Shi, Muennighoff, Wang, Liu, Shi, Siegel, Tang et~al.}]{su2024bright}
Hongjin Su, Howard Yen, Mengzhou Xia, Weijia Shi, Niklas Muennighoff, Han-yu Wang, Haisu Liu, Quan Shi, Zachary~S Siegel, Michael Tang, et~al. 2024.
\newblock Bright: A realistic and challenging benchmark for reasoning-intensive retrieval.
\newblock \emph{arXiv preprint arXiv:2407.12883}.

\bibitem[{Wang et~al.(2023)Wang, Yang, Huang, Yang, Majumder, and Wei}]{wang2023improving}
Liang Wang, Nan Yang, Xiaolong Huang, Linjun Yang, Rangan Majumder, and Furu Wei. 2023.
\newblock Improving text embeddings with large language models.
\newblock \emph{arXiv preprint arXiv:2401.00368}.

\bibitem[{Wang et~al.(2024{\natexlab{a}})Wang, Yang, Huang, Yang, Majumder, and Wei}]{wang-etal-2024-improving-text}
Liang Wang, Nan Yang, Xiaolong Huang, Linjun Yang, Rangan Majumder, and Furu Wei. 2024{\natexlab{a}}.
\newblock \href {https://doi.org/10.18653/v1/2024.acl-long.642} {Improving text embeddings with large language models}.
\newblock In \emph{Proceedings of the 62nd Annual Meeting of the Association for Computational Linguistics (Volume 1: Long Papers)}, pages 11897--11916, Bangkok, Thailand. Association for Computational Linguistics.

\bibitem[{Wang et~al.(2024{\natexlab{b}})Wang, Wang, Cao, Wang, Paturi, and Bergen}]{wang2024bircobenchmarkinformationretrieval}
Xiaoyue Wang, Jianyou Wang, Weili Cao, Kaicheng Wang, Ramamohan Paturi, and Leon Bergen. 2024{\natexlab{b}}.
\newblock \href {https://arxiv.org/abs/2402.14151} {Birco: A benchmark of information retrieval tasks with complex objectives}.
\newblock \emph{Preprint}, arXiv:2402.14151.

\bibitem[{Wang et~al.(2024{\natexlab{c}})Wang, Zhang, and Wang}]{wang2024generateddatahelpcontrastive}
Yifei Wang, Jizhe Zhang, and Yisen Wang. 2024{\natexlab{c}}.
\newblock \href {https://arxiv.org/abs/2403.12448} {Do generated data always help contrastive learning?}
\newblock \emph{Preprint}, arXiv:2403.12448.

\bibitem[{Weller et~al.(2024{\natexlab{a}})Weller, Chang, MacAvaney, Lo, Cohan, Van~Durme, Lawrie, and Soldaini}]{weller2024followir}
Orion Weller, Benjamin Chang, Sean MacAvaney, Kyle Lo, Arman Cohan, Benjamin Van~Durme, Dawn Lawrie, and Luca Soldaini. 2024{\natexlab{a}}.
\newblock Followir: Evaluating and teaching information retrieval models to follow instructions.
\newblock \emph{arXiv preprint arXiv:2403.15246}.

\bibitem[{Weller et~al.(2024{\natexlab{b}})Weller, Lawrie, and Durme}]{weller2024nevirnegationneuralinformation}
Orion Weller, Dawn Lawrie, and Benjamin~Van Durme. 2024{\natexlab{b}}.
\newblock \href {https://arxiv.org/abs/2305.07614} {Nevir: Negation in neural information retrieval}.
\newblock \emph{Preprint}, arXiv:2305.07614.

\bibitem[{Weller et~al.(2024{\natexlab{c}})Weller, Van~Durme, Lawrie, Paranjape, Zhang, and Hessel}]{weller2024promptriever}
Orion Weller, Benjamin Van~Durme, Dawn Lawrie, Ashwin Paranjape, Yuhao Zhang, and Jack Hessel. 2024{\natexlab{c}}.
\newblock Promptriever: Instruction-trained retrievers can be prompted like language models.
\newblock \emph{arXiv preprint arXiv:2409.11136}.

\bibitem[{Xiao et~al.(2023)Xiao, Liu, Zhang, and Muennighoff}]{bge_embedding}
Shitao Xiao, Zheng Liu, Peitian Zhang, and Niklas Muennighoff. 2023.
\newblock \href {https://arxiv.org/abs/2309.07597} {C-pack: Packaged resources to advance general chinese embedding}.
\newblock \emph{Preprint}, arXiv:2309.07597.

\bibitem[{Zhan et~al.(2021)Zhan, Mao, Liu, Guo, Zhang, and Ma}]{zhan2021optimizing}
Jingtao Zhan, Jiaxin Mao, Yiqun Liu, Jiafeng Guo, Min Zhang, and Shaoping Ma. 2021.
\newblock Optimizing dense retrieval model training with hard negatives.
\newblock In \emph{Proceedings of the 44th International ACM SIGIR Conference on Research and Development in Information Retrieval}, pages 1503--1512.

\bibitem[{Zhang et~al.(2024{\natexlab{a}})Zhang, Zhang, Wu, Pei, Ren, de~Rijke, Chen, and Ren}]{zhang2024excluir}
Wenhao Zhang, Mengqi Zhang, Shiguang Wu, Jiahuan Pei, Zhaochun Ren, Maarten de~Rijke, Zhumin Chen, and Pengjie Ren. 2024{\natexlab{a}}.
\newblock Excluir: Exclusionary neural information retrieval.
\newblock \emph{arXiv preprint arXiv:2404.17288}.

\bibitem[{Zhang et~al.(2024{\natexlab{b}})Zhang, Yu, Fu, Huang, and Li}]{zhang2024iopo}
Xinghua Zhang, Haiyang Yu, Cheng Fu, Fei Huang, and Yongbin Li. 2024{\natexlab{b}}.
\newblock Iopo: Empowering llms with complex instruction following via input-output preference optimization.
\newblock \emph{arXiv preprint arXiv:2411.06208}.

\bibitem[{Zhang et~al.(2025)Zhang, Yang, Prenkaj, and Kasneci}]{zhang2025featuresdeserveattentiongraphguided}
Zheyu Zhang, Shuo Yang, Bardh Prenkaj, and Gjergji Kasneci. 2025.
\newblock \href {https://arxiv.org/abs/2507.18504} {Not all features deserve attention: Graph-guided dependency learning for tabular data generation with language models}.
\newblock \emph{Preprint}, arXiv:2507.18504.

\bibitem[{Zhang et~al.(2024{\natexlab{c}})Zhang, Zhu, Zhou, Qi, Zhang, and Li}]{zhang2024boolquestionsdoesdenseretrieval}
Zongmeng Zhang, Jinhua Zhu, Wengang Zhou, Xiang Qi, Peng Zhang, and Houqiang Li. 2024{\natexlab{c}}.
\newblock \href {https://arxiv.org/abs/2411.12235} {Boolquestions: Does dense retrieval understand boolean logic in language?}
\newblock \emph{Preprint}, arXiv:2411.12235.

\bibitem[{Zhu et~al.(2025)Zhu, Wan, Steck, Liang, Feng, Kallus, and Li}]{zhu2025collaborative}
Yaochen Zhu, Chao Wan, Harald Steck, Dawen Liang, Yesu Feng, Nathan Kallus, and Jundong Li. 2025.
\newblock Collaborative retrieval for large language model-based conversational recommender systems.
\newblock In \emph{Proceedings of the ACM on Web Conference 2025}, pages 3323--3334.

\bibitem[{Zhu et~al.(2023)Zhu, Yuan, Wang, Liu, Liu, Deng, Chen, Liu, Dou, and Wen}]{zhu2023large}
Yutao Zhu, Huaying Yuan, Shuting Wang, Jiongnan Liu, Wenhan Liu, Chenlong Deng, Haonan Chen, Zheng Liu, Zhicheng Dou, and Ji-Rong Wen. 2023.
\newblock Large language models for information retrieval: A survey.
\newblock \emph{arXiv preprint arXiv:2308.07107}.

\end{thebibliography}

\newpage
\appendix
\onecolumn 

\section{Details of Models}
\label{sec:details of models}
For each model used in this paper, Table \ref{tab:details of models} provides details on model size, architecture, maximum input context length, and whether instructions is included. The GPT-4o model utilized for dataset generation and ranking in this work was the "2024-07-01-preview" version.

\begin{table*}[htbp]
  \centering
  \resizebox{\textwidth}{!}{
    \begin{tabular}{cccccc}
    \toprule
    Model & Size  & Architecture & Instruction & Max length & Pooling Method\\
    \midrule
    \multicolumn{6}{c}{Sparse Retriever} \\
    \midrule
    \rowcolor[rgb]{ 1,  .953,  .949} BM25~\citep{robertson2009bm25}  & N/A   & Sparse & No    & N/A & N/A \\
    \midrule
    \multicolumn{6}{c}{Dense Retriever} \\
    \midrule
    \rowcolor[rgb]{ .949,  .949,  1} jina-embeddings-v3~\citep{sturua2024jina} & 572M  & Encoder & No    & 4K & Mean\\

    \rowcolor[rgb]{ .949,  .949,  1} gte-large-en-v1.5~\citep{li2023towards} & 434M  & Encoder & No    & 8K & \texttt{<GLS>} \\

    \rowcolor[rgb]{ .949,  .949,  1} NV-Embed-v2~\citep{lee2024nv} & 7.8B  & Decoder & Yes   & 32K & Latent Attention Layer \\

    \rowcolor[rgb]{ .949,  .949,  1} bge-en-icl~\citep{li2024makingtextembeddersfewshot} & 7.1B  & Decoder & Yes   & 32K & \texttt{<EOS>}\\

    \rowcolor[rgb]{ .949,  .949,  1} gte-Qwen2-7B-instruct~\citep{li2023towards} & 7.6B  & Decoder & Yes   & 131K & \texttt{<EOS>}\\

    \rowcolor[rgb]{ .949,  .949,  1} gte-Qwen2-1.5B-instruct~\citep{li2023towards} & 1.5B  & Decoder & Yes   & 131K & \texttt{<EOS>}\\

    \rowcolor[rgb]{ .949,  .949,  1} e5-mistral-7b-instruct~\citep{wang-etal-2024-improving-text} & 7.1B  & Decoder & Yes   & 32K & \texttt{<EOS>}\\

    \rowcolor[rgb]{ .949,  .949,  1} GritLM-7B~\citep{muennighoff2024generative} & 7.2B  & Decoder & Yes   & 4K & Mean\\

    \rowcolor[rgb]{ .949,  .949,  1} LLM2Vec~\citep{behnamghader2024llm2veclargelanguagemodels} & 7.5B  & Decoder & Yes   & 8K & Mean\\
    \midrule
    \multicolumn{6}{c}{Fine-tuned Reranker} \\
    \midrule
    \rowcolor[rgb]{ .949,  1,  .949} bge-reranker-v2-m3~\citep{chen2024bge} & 568M  & Cross-Encoder & No    & 8k & N/A\\

    \rowcolor[rgb]{ .949,  1,  .949} bge-reranker-v2-gemma~\citep{chen2024bge} & 2.5B  & Decoder & Yes   & 8K & N/A\\
    \rowcolor[rgb]{ .949,  1,  .949} followIR~\citep{weller2024followir} & 7.2B  & Decoder & Yes   & 4K & N/A\\
    \rowcolor[rgb]{ .949,  1,  .949} RankZephyr~\citep{pradeep2023rankzephyr} & 7B  & Decoder & Yes   & 32K & N/A\\
    \midrule 
    \multicolumn{6}{c}{Zero-shot LLM for Ranking} \\
    \midrule
    \rowcolor[rgb]{  .949,  .949,  .949} 
    GPT-4o~\citep{OpenAI2024helloGPT4o} & N/A  & Decoder & Yes   & 128K & N/A\\
    \bottomrule
    \end{tabular}%
    }
    \caption{\textbf{Details of models used in experiments. }We list the number of parameters of each model except the sparse model (BM25). Regarding the model architecture, we distinguish between sparse models, dense models, and rerankers. Dense models are further classified as Encoders or Decoders. Rerankers are categorized into Cross Encoders and Decoders (LLM-based generative relevance scoring). The Instruction column indicates whether instructions are included in the retrieval process. Max length denotes the maximum input length used for each model in the experiments. The Pooling Method represents the approach used by the model to obtain embeddings.}
  \label{tab:details of models}%
\end{table*}%

For Dense Retrieval models that require instructions (NV-Embed-v2, bge-en-icl, gte-Qwen2-7B-instruct, gte-Qwen2-1.5B-instruct, e5-mistral-7b-instruct, GritLM-7B, and LLM2Vec), we use the following instruction:

\textit{“Given a {domain} retrieval query, retrieve documents that meet the specified conditions.”}

For LLM-based rerankers (bge-reranker-v2-gemma and followIR), we adopt the model’s default prompt. For example, bge-reranker-v2-gemma uses the following prompt:

\textit{“Given a query A and a passage B, determine whether the passage contains an answer to the query by providing a prediction of either ‘Yes’ or ‘No’.”}

For models that do not require instructions, we directly input the query and document, such as jina-embeddings-v3, gte-large-en-v1.5, and bge-reranker-v2-m3.

\section{Prompt Templates For Constructing \textbf{\textsc{Multiconir Dataset}}}
\label{sec:prompt}
Table \ref{tab:task1_prompt}, \ref{tab:task2_prompt}, and \ref{tab:task3_prompt} present the prompts used in Steps 1 to 3 for constructing our \textbf{\textsc{Multiconir}} dataset. 

For placeholders,
\{\textit{domain}\} $\in$ \{\textit{People, Books, Movies, Medical\ Case, Legal\ Document\}}. 
\{\textit{domain\_features}\} specifies key attributes within a particular domain. In the medical case domain, \{\textit{domain\_features}\} $\in$ \{ \textit{patient symptoms, clinical diagnosis, drug allergies, family medical history, surgical details, postoperative outcomes, hospitalization duration, recovery status.}\}
In the legal document domain, \{\textit{domain\_features}\} $\in$ \{ \textit{case type, involved parties, court ruling, legal provisions, evidence summary, defense strategy.} \}  
In the movies domain, \{\textit{domain\_features}\} $\in$ \{ \textit{ summary, lead actors, release date, release area, genre, detailed plots.} \}  
In the books domain, \{\textit{domain\_features}\} $\in$ \{ \textit{ author, publication year, genre, main content, detailed plots.} \}  
In the people domain, \{\textit{domain\_features}\} $\in$ \{ \textit{profession, nationality, notable achievements, social impact, related events.} \}  
\begin{table}[!ht]
\centering
\begin{tabular}{p{3cm}p{12cm}} 
\hline
\textbf{Task} & \textbf{Prompt} \\ 
\hline
\textbf{Step 1: Condition Sentence Extraction} &
I will provide you a document of \{domain\}, you should extract ten detailed sentences that represent the key conditions the document satisfies. \newline 

Please adhere to the following guidelines: \newline 
- Extract fine-grained condition-related sentences relevant to \{domain\}, such as \{domain\_features\}. \newline
- Do not paraphrase; use the original sentences from the document. \newline
- Ensure each sentence is semantically intact and not conflict with the context. \newline
% - Exclude sentences containing 'keywords: nan' or 'none'. \newline
- Format the output as an array, e.g., [“sentence1”, “sentence2”, ..., “sentence10”]. \newline

Here is the document: \{domain\_document\}. \newline 
Return array only. \\ 
\hline
\end{tabular}
\caption{Prompt for GPT-4o to extract condition sentence (Step 1).}
\label{tab:task1_prompt}
\end{table}

\begin{table}[!ht]
\centering
\begin{tabular}{p{3cm}p{12cm}} 
\hline
\textbf{Task} & \textbf{Prompt} \\ 
\hline
\textbf{Step 2: Query Generation (Instruction-style)} &
I will provide you \{num\} condition-related sentences; formulate a retrieval query for me.\newline

Here are a few examples for reference: \newline
- \{Instruction-style example 1\} \newline
- \{Instruction-style example 2\} \newline

Please adhere to the following guidelines: \newline 
- Each sentence represents a condition; with \{num\} sentences, the number of conditions is \{num\}. \newline
- The query should be instruction-style, explicitly listing all conditions.
- Each condition should be around 10 words. \newline
- Make conditions concise, summarizing each sentence. \newline
- You can paraphrase and modify keywords while maintaining meaning. \newline

Here are the sentences: \{info\}. \newline 
Return one query only. Do not include extra information. \\ 

\hline
\textbf{Step 2: Query Generation (Descriptive-style)} &
I will provide you \{num\} condition-related sentences; formulate a retrieval query for me.\newline

Here are a few examples for reference: \newline
- \{Descriptive-style example 1\} \newline
- \{Descriptive-style example 2\} \newline

Please adhere to the following guidelines: \newline 
- Each sentence represents a condition; with \{num\} sentences, the number of conditions is \{num\}. \newline
- The query should be descriptive-style, integrating and describing all conditions in natural language.\newline
- Each condition should be around 10 words. \newline
- Make conditions concise, summarizing each sentence. \newline
- You can paraphrase and modify keywords while maintaining meaning. \newline

Here are the sentences: \{info\}. \newline 
Return one query only. Do not include extra information. \\ 
\hline
\end{tabular}
\caption{Prompt for GPT-4o to generate queries with varying conditions (Step 2).}
\label{tab:task2_prompt}
\end{table}

\begin{table}[!ht]
\centering
\begin{tabular}{p{3cm}p{12cm}} 
\hline
\textbf{Task} & \textbf{Prompt} \\ 
\hline
\textbf{Step 3: Hard Negative Sentence Making (For Books, Movies, Medical Case, and Legal Document Datasets)} &
I will provide you one query and one sentence, generate a modified sentence for me. \newline

Here are a few examples for reference: \newline
Query: - \{query\} \newline
Sentence: - \{condition sentence\} \newline
Modified: - \{hard negative sentence\} \newline

Please adhere to the following guidelines: \newline 
- Modify the sentence so that its meaning no longer aligns with the query. \newline
- Keep key terms unchanged. \newline
- Ensure the new sentence is semantically different from the original. \newline

Here is the query: \{query\}. \newline
Here is the Sentence: \{information\}. \newline
Return only the modified sentence. \\ 
\hline

\hline
\textbf{Step 3: Hard Negative Sentence Making (For People Dataset)} &
I will provide you one query and one sentence, generate a modified sentence for me. \newline

Here are a few examples for reference: \newline
Query: - Who is the American artist that went to RISD? \newline
Sentence: - He went to RISD for graduate school. \newline
Modified: - He went to ACCA for graduate school, but his sister went to RISD. \newline

Please adhere to the following guidelines: \newline 
- Modify the sentence so that its meaning no longer aligns with the query. \newline
- Keep key terms unchanged, but introduce dummy information to mislead the retrieval model. For example, if the original sentence states, “He went to RISD for graduate school,” you can modify it to, “He went to ACCA for graduate school, but his sister went to RISD,” where the key term (RISD) remains but is assigned to an irrelevant entity (his sister). \newline
- Ensure the new sentence is semantically different from the original by using different wording and synonymous substitution. \newline
- The changed sentence should prevent the query from retrieving it as relevant information. \newline

Here is the query: \{query\}. \newline
Here is the sentence: \{information\}. \newline
Return only the modified sentence. \\ 
\hline
\end{tabular}
\caption{Prompt for GPT-4o to modify the condition sentence to hard negative sentence (Step 3).}
\label{tab:task3_prompt}
\end{table}

\section{Benchmark Quality Evaluation}
\label{sec:dataset quality}

To guarantee the reliability of \textsc{MultiConIR}, we applied a two–stage audit that examines both \emph{query validity} and \emph{document–label validity}.  Figure~\ref{fig:quality} gives a visual outline; full numbers appear in Table~\ref{tab:domain_sources}.

\paragraph{Query validity.}
From each of the five domains (\textit{People}, \textit{Books}, \textit{Movies}, \textit{Medical}, \textit{Legal}) we randomly sampled 100 multi-condition queries, yielding a 500-item evaluation set.  
Ten trained graduate annotators independently rated every query for \emph{linguistic naturalness}, \emph{precision of constraints}, and \emph{contextual plausibility} (i.e., whether logically plausible).  
Inter-annotator agreement reached \(93.7\%\) with Fleiss’~\(\kappa = 0.84\), indicating near-perfect consensus that the automatically generated queries resemble genuine information needs.

\paragraph{Document–label validity.}
\begin{enumerate}[leftmargin=1.6em, itemsep=2pt]
    \item \textbf{LLM filtering.}  
    We applied \textit{GPT-4o} to the \emph{entire} corpus.  
    For every positive document \(d^{+}\) the model verified that all ten conditions in \(C(d)\) were satisfied;  
    for each hard-negative document \(d^{-}_{k}\) it checked that \emph{exactly} \(k\!-\!1\) conditions held.  
    Instances failing these criteria (false positives or false negatives) were discarded, reducing domain sizes to:  
    People~(420), Books~(482), Movies~(500), Medical~(479), Legal~(426).
    
    \item \textbf{Human spot-check.}  
    To confirm the LLM filter, we randomly drew another 100 document–query pairs per domain (500 in total).  
    Two independent annotators judged whether the labelled number of satisfied conditions was correct; disagreements were resolved by a third adjudicator.  
    The residual error rate was \(2.4\%\), implying that the automatic filter removed the vast majority of mis-labelled items.
\end{enumerate}

\section{Discussion: \emph{Win Rate} vs.\ Traditional IR Metrics}
\label{sec:metric_discussion}

\paragraph{Why introduce \textbf{Win Rate}?}
\textsc{MultiConIR} poses \emph{multi-condition} queries for which models must sense fine-grained semantic differences.  
We focus on two abilities:  
\textbf{(i)} discriminating the \emph{positive} document from a hard negative as the number of query conditions grows (\textbf{Task 1});  
\textbf{(ii)} preserving a \emph{monotonic ordering} in which a document satisfying \(k\) conditions outranks one satisfying \(k\!-\!1\) under the \emph{same} query (\textbf{Task 2}).

\paragraph{Limitations of conventional metrics.}
\begin{itemize}[leftmargin=1.4em]
    \item \textbf{Precision@1} coincides with Win Rate in Task 1 (one positive vs.\ one HN) but, in Task 2, observes only the \emph{top} result and ignores the intended hierarchy
          \(d^{+}\succ\mathrm{HN}_1\succ\cdots\succ\mathrm{HN}_{10}\).
    \item \textbf{NDCG@\(k\)} introduces graded relevance but still weights absolute rank more than pairwise consistency, thus blurring step-wise violations of the monotonic order.
    \item \textbf{Recall} proved even less informative in early pilot runs dominated by easy negatives: high recall was achievable without respecting the semantic precision that  
          \textsc{MultiConIR} is designed to test.
\end{itemize}

\paragraph{How Win Rate fills the gap.}
Win Rate computes the proportion of pairwise comparisons in which a document that fulfils more conditions is ranked above one that fulfils fewer.  
Hence, it \emph{matches Precision@1} in the degenerate Task 1 case, yet remains sensitive to every local inversion in the graded Task 2 ladder, offering a sharper lens on a model’s ability to capture incremental semantic constraints.

\section{Examples Of The \textbf{\textsc{MultiConIR}} Dataset}
\label{sec:dataset}

% Table \ref{tab:domain_sources} presents the five domains and their source data used for constructing our \textbf{\textit{MultiConIR}} dataset. 
Tables \ref{tab:book_ex}, \ref{tab:movie_ex}, \ref{tab:people_ex}, \ref{tab:medical_ex}, and \ref{tab:legal_ex} illustrate examples from their respective domains.
% \begin{table*}[htbp]
%   \centering

%   \resizebox{=0.75\textwidth}{!}
%   {%
%     \begin{tabular}{cccccc}
%     \hline
%     Domain & Records  & Source Data \\
%     \hline
%     People & 420   & People Wikipedia Dat\citep{mahajan_people_wikipedia_data} \\
%     Books & 482    & Books Dataset \citep{rustamov_books_dataset} \\
%     Movies & 500   & Wikipedia Movie Plots \citep{robischon_wikipedia_movie_plots} \\
%     Medical Case & 479    & Medical Cases \citep{hpe2023medicalcases} \\
%     Legal Document & 426    & LexGLUE \citep{chalkidis-etal-2022-lexglue} \\
%     \hline
%     \end{tabular}%
%     }
%     \caption{Domain-specific Source Data References. The Records column indicates the number of data entries for each domain in the MultiConIR dataset.}
%     \label{tab:domain_sources}%
% \end{table*}%

% people
\begin{table*}[!ht]
\centering
\begin{tabular}{p{4.5cm}p{5cm}p{5cm}}
\hline
\textbf{Query 3} & \textbf{Positive} & \textbf{HN 1} \\ 
\hline
Find a notable individual who meets these criteria:
\setlist[enumerate]{itemsep=0.12em, topsep=0.5em}
\begin{enumerate}
    \item Studies plasma melatonin to assess biological rhythm disorders.
    \item Identified 25-hour circadian rhythms in totally blind individuals.
    \item \textbf{Worked at NIMH in Bethesda, Maryland before 1981.}
    % \item Collaborated closely with Robert Sack in Oregon.
    % \item Researches chronobiologic sleep and mood disorders.
    % \item Graduated from University of Chicago in nineteen seventies with MD and PhD.
    % \item Had around 90 publications on PubMed by 2005.
    % \item Is a full professor and vice-chair of Psychiatry at OHSU.
    % \item Focuses on bright light exposure and melatonin treatments.
    % \item \textbf{Studies disorders like winter depression, jet lag, and shift work maladaptation.}
\end{enumerate}

&

Relying on a very precise assay for plasma melatonin, a hormone that has a clearly defined 24-hour pattern of secretion, biological rhythm disorders can be assessed and their treatment can be monitored. Totally blind individuals have 25-hour circadian rhythms, drifting an hour later each day unless they take a melatonin capsule at a certain time every day. \textbf{Prior to moving to Oregon in 1981, Lewy was at the National Institute of Mental Health (NIMH) in Bethesda, Maryland, working with senior colleague Thomas Wehr.} In Oregon, he has worked closely with Robert L. Sack. He describes his research as follows: 'My laboratory studies chronobiologic sleep and mood disorders.' Alfred J. Lewy, aka Sandy Lewy, graduated from University of Chicago in 1973 after studying psychiatry, pharmacology, and ophthalmology. As of December 2005, he had 94 publications available on PubMed. He is a full professor and vice-chair of the Department of Psychiatry at OHSU (Oregon Health Science University) and holds an MD and PhD. Current research is focused on developing bright light exposure and melatonin administration as treatment modalities for these disorders. These disorders include winter depression, jet lag, maladaptation to shift work, and certain types of sleep disturbances.

&
Relying on a very precise assay for plasma melatonin, a hormone that has a clearly defined 24-hour pattern of secretion, biological rhythm disorders can be assessed and their treatment can be monitored. Totally blind individuals have 25-hour circadian rhythms, drifting an hour later each day unless they take a melatonin capsule at a certain time every day. \textcolor{red}{After to moving to Oregon in 1981, Lewy was at the National Institute of Mental Health (NIMH) in Bethesda, Maryland, working with senior colleague Thomas Wehr}. In Oregon, he has worked closely with Robert L. Sack. He describes his research as follows: 'My laboratory studies chronobiologic sleep and mood disorders.' Alfred J. Lewy, aka Sandy Lewy, graduated from University of Chicago in 1973 after studying psychiatry, pharmacology, and ophthalmology. As of December 2005, he had 94 publications available on PubMed. He is a full professor and vice-chair of the Department of Psychiatry at OHSU (Oregon Health Science University) and holds an MD and PhD. Current research is focused on developing bright light exposure and melatonin administration as treatment modalities for these disorders. These disorders include winter depression, jet lag, maladaptation to shift work, and certain types of sleep disturbances.\\
\hline
\end{tabular}
\caption{An example in domain of People}
\label{tab:people_ex}
\end{table*}

% book
\begin{table*}[!ht]
\centering
\begin{tabular}{p{4.5cm}p{5cm}p{5cm}}
\hline
\textbf{Query 5} & \textbf{Positive} & \textbf{HN 1} \\ 
\hline
Find a notable individual who meets these criteria:
\setlist[enumerate]{itemsep=0.12em, topsep=0.5em}
\begin{enumerate}
    \item Details Bernie Madoff's \$65B Ponzi collapse.
    \item Covers the impact on national media.
    \item Investigates Madoff's history of fraud.
    \item Offers deep insight into Madoff's family.
    \item \textbf{Contains exclusive news and material.}
    % \item Includes Bernie's Little Black Book.
    % \item Features Madoff's past calendars.
    % \item Provides details on the origins of his scam.
    % \item Includes cooperation from Eleanor Squillari.
    % \item \textbf{Reveals SEC connections to Madoff.}
\end{enumerate}
&

The collapse of Bernie Madoff's Ponzi scheme led to the instant evaporation of \$65 billion of wealth. The effects of Madoff's brazen fraud were felt most closely in New York and Palm Beach but the story was, and continues to be, front page news across the country. Brian Ross and his team of investigators shed an unyielding light onto Madoff's scheme--how he got started, how he succeed for so long, who helped him, and who shielded him from early investigations. This is an incisive and voyeuristic look into this first family of financial crime. \textbf{The Madoff Chronicles includes a vast array of news and material that readers won't find anywhere else.} Contains a reproduction of Bernie's Little Black Book. Ross has also secured Madoff's calendar for the past three years and other never-before-seen documents from inside the Madoff empire, straight from his desk. Read key details of how Madoff carried out his scam and the revelation that he began the fraud from almost the first day, in the 1960s. Extensive cooperation by Madoff's personal assistant, Eleanor Squillari. Contains incriminating connections between Madoff and certain members of the SEC.

&
The collapse of Bernie Madoff's Ponzi scheme led to the instant evaporation of \$65 billion of wealth. The effects of Madoff's brazen fraud were felt most closely in New York and Palm Beach but the story was, and continues to be, front page news across the country. Brian Ross and his team of investigators shed an unyielding light onto Madoff's scheme--how he got started, how he succeed for so long, who helped him, and who shielded him from early investigations. This is an incisive and voyeuristic look into this first family of financial crime. \textcolor{red}{The Madoff Chronicles includes a vast array of news and material.} Contains a reproduction of Bernie's Little Black Book. Ross has also secured Madoff's calendar for the past three years and other never-before-seen documents from inside the Madoff empire, straight from his desk. Read key details of how Madoff carried out his scam and the revelation that he began the fraud from almost the first day, in the 1960s. Extensive cooperation by Madoff's personal assistant, Eleanor Squillari. Contains incriminating connections between Madoff and certain members of the SEC.\\
\hline
\end{tabular}
\caption{An example in domain of Book}
\label{tab:book_ex}
\end{table*}

% movie

\begin{table*}[!ht]
\centering
\begin{tabular}{p{4.5cm}p{5cm}p{5cm}}
\hline
\textbf{Query 8} & \textbf{Positive} & \textbf{HN 1} \\ 
\hline

Find a movie that matches all conditions: 
\setlist[enumerate]{itemsep=0.12em, topsep=0.5em}
\begin{enumerate}
    \item Originated from American.
    \item Plot: Mrs. Lowe and Black were lovers.
    \item Plot: Terry carelessly spends money.
    \item Plot: Terry promoted to ship foreman.
    \item Cast: Mary Miles Minter, Allan Forrest.
    \item Plot: Julia Deep works behind exchange desk.
    \item Director: Lloyd Ingraham.
    \item \textbf{Plot: Terry's spending takes a toll.}
    % \item Plot: Lottie sees Terry and Julia at park.
    % \item Release Year: 1918.
\end{enumerate}
&

Origin/Ethnicity: American \newline
Meanwhile, it is revealed Mrs. Lowe and Black were once lovers. He is spending his money carelessly and doesn't put any time in paying the bills, much to the dislike of the department store owner Timothy Black. Soon, Terry is promoted to a foreman on a ship.\newline
Cast: Mary Miles Minter, Allan Forrest Julia Deep is a young woman working behind the exchange desk at a department store. \newline
Director: Lloyd Ingraham \newline
\textbf{After a while, Terry's money spending takes its toll.} Lottie gets distracted and does not notice Terry and Julia at the park. \newline
Release Year: 1918

&
Origin/Ethnicity: American \newline
Meanwhile, it is revealed Mrs. Lowe and Black were once lovers. He is spending his money carelessly and doesn't put any time in paying the bills, much to the dislike of the department store owner Timothy Black. Soon, Terry is promoted to a foreman on a ship. \newline
Cast: Mary Miles Minter, Allan Forrest Julia Deep is a young woman working behind the exchange desk at a department store. \newline
Director: Lloyd Ingraham        \newline
\textbf{\color{red}{Eventually, Terry's frugality leads to financial growth.}} Lottie gets distracted and does not notice Terry and Julia at the park.\newline
Release Year: 1918 \\
\hline

\end{tabular}
\caption{An example in domain of Movie}
\label{tab:movie_ex}
\end{table*}

% medical case

\begin{table*}[!ht]
\centering
\begin{tabular}{p{4.5cm}p{5cm}p{5cm}}
\hline
\textbf{Query 7} & \textbf{Positive} & \textbf{HN 1} \\ 
\hline
Find a case where the patient:
\setlist[enumerate]{itemsep=0em, topsep=0.5em}
\begin{enumerate}
    \item Underwent ascending aortic arch angiogram.
    \item Had left common carotid artery angiogram.
    \item Received right common carotid artery angiogram.
    \item Undergone left subclavian artery angiogram.
    \item Had right iliac angiogram with runoff.
    \item Performed bilateral cerebral angiograms.
    \item \textbf{Experienced TIA and moderate carotid stenosis.}
    % \item Had 400 ml blood loss.
    % \item Provided informed consent for the procedure.
    % \item \textbf{Received a 6-French sheath in the right femoral artery.}
\end{enumerate}
&

PROCEDURE PERFORMED: \newline
1.  Selective ascending aortic arch angiogram. 2.  Selective left common carotid artery angiogram. 3.  Selective right common carotid artery angiogram. 4.  Selective left subclavian artery angiogram. 5.  Right iliac angio with runoff. 6.  Bilateral cerebral angiograms were performed as well via right and left common carotid artery injections. \newline \textbf{INDICATIONS FOR PROCEDURE: TIA, aortic stenosis, postoperative procedure. Moderate carotid artery stenosis.} \newline ESTIMATED BLOOD LOSS: 400 ml. \newline
After obtaining informed consent, the patient was brought to the cardiac catheterization suite in postabsorptive and nonsedated state. Using modified Seldinger technique, a 6-French sheath was placed into the right common femoral artery and vein without complication.

&
PROCEDURE PERFORMED: \newline
1.  Selective ascending aortic arch angiogram. 2.  Selective left common carotid artery angiogram. 3.  Selective right common carotid artery angiogram. 4.  Selective left subclavian artery angiogram. 5.  Right iliac angio with runoff. 6.  Bilateral cerebral angiograms were performed as well via right and left common carotid artery injections. \newline \textcolor{red}{INDICATIONS FOR PROCEDURE: TIA, aortic stenosis, postoperative procedure. Severe carotid artery stenosis.} \newline ESTIMATED BLOOD LOSS: 400 ml. \newline 
After obtaining informed consent, the patient was brought to the cardiac catheterization suite in postabsorptive and nonsedated state. A 6-French sheath was used in the left femoral artery and vein with minor complications, employing the modified Seldinger technique.\\
\hline

\end{tabular}
\caption{An example in domain of Medical Case}
\label{tab:medical_ex}
\end{table*}

% legal document

\begin{table*}[!ht]
\centering
\begin{tabular}{p{3cm}p{6cm}p{6cm}}
\hline
\textbf{Query 10} & \textbf{Positive} & \textbf{HN 1} \\ 
\hline

Find a case where:  \newline
1. Michigan Legislature enacted a statute in 1987. \newline
2. Petitioners challenged the statute under Contract Clause and Due Process Clause. \newline
3. The statute affected workers injured before March 31, 1982. \newline
4. Petitioners argued a 1981 law allowed reduction of workers' compensation benefits. \newline
5. The Michigan Supreme Court accepted petitioners' interpretation in 1985. \newline
6. Legislature introduced a bill to overturn the court's decision. \newline
7. House Bill 5084 was introduced in October 1985. \newline
8. The bill became law on May 14, 1987. \newline
9. Petitioners were ordered to refund nearly \$25 million. \newline
10. \textbf{Michigan Supreme Court upheld the statute for lacking vested rights and rational purpose.}
&

In 1987, the Michigan Legislature enacted a statute that had the effect of requiring petitioners General Motors Corporation (GM) and Ford Motor Company (Ford) to repay workers' compensation benefits GM and Ford had withheld in reliance on a 1981 workers' compensation statute. Petitioners challenge the provision of the statute mandating these retroactive payments on the ground that it violates the Contract Clause and the Due Process Clause of the Federal Constitution. The benefit coordination provision did not specify whether it was to be applied to workers injured before its effective date, March 31, 1982. Petitioners took the position that the 1981 law allowed them to reduce workers' compensation benefits to workers injured before March 31, 1982, who were receiving benefits from other sources. In 1985, petitioners' interpretation was accepted by the Michigan Supreme Court. Chambers v. General Motors Corp., decided together with Franks v. White Pine Copper Div., Copper Range Co., 422 Mich. 636, 375 N.W.2d 715. The Michigan Legislature responded almost immediately by introducing legislation to overturn the court's decision. On October 16, 1985, before the Michigan Supreme Court had ruled on the motion for rehearing in Chambers, House Bill 5084 was introduced. The amended Senate bill passed into law on May 14, 1987. 1987 Mich.Pub.Acts No. 28. As a result of the 1987 statute, petitioners were ordered to refund nearly \$25 million to disabled employees. \textbf{The Michigan Supreme Court upheld the statute against these challenges, on the ground that the employers had no vested rights in coordination for Contract Clause purposes, and that the retroactive provisions furthered a rational legislative purpose.} 436 Mich. 515, 462 N.W.2d 555 (1990).

&
In 1987, the Michigan Legislature enacted a statute that had the effect of requiring petitioners General Motors Corporation (GM) and Ford Motor Company (Ford) to repay workers' compensation benefits GM and Ford had withheld in reliance on a 1981 workers' compensation statute. Petitioners challenge the provision of the statute mandating these retroactive payments on the ground that it violates the Contract Clause and the Due Process Clause of the Federal Constitution. The benefit coordination provision did not specify whether it was to be applied to workers injured before its effective date, March 31, 1982. Petitioners took the position that the 1981 law allowed them to reduce workers' compensation benefits to workers injured before March 31, 1982, who were receiving benefits from other sources. In 1985, petitioners' interpretation was accepted by the Michigan Supreme Court. Chambers v. General Motors Corp., decided together with Franks v. White Pine Copper Div., Copper Range Co., 422 Mich. 636, 375 N.W.2d 715. The Michigan Legislature responded almost immediately by introducing legislation to overturn the court's decision. On October 16, 1985, before the Michigan Supreme Court had ruled on the motion for rehearing in Chambers, House Bill 5084 was introduced. The amended Senate bill passed into law on May 14, 1987. 1987 Mich.Pub.Acts No. 28. As a result of the 1987 statute, petitioners were ordered to refund nearly \$25 million to disabled employees. \textbf{\color{red}{The Michigan Supreme Court found the statute invalid on the grounds that the retroactive provisions did not further a rational legislative purpose and that the employers had vested rights in coordination for Contract Clause purposes.}} 436 Mich. 515, 462 N.W.2d 555 (1990).\\
\hline
\end{tabular}
\caption{An example in domain of Legal Document}
\label{tab:legal_ex}
\end{table*}

\section{Complete Results}
\label{sec:complete results}

\subsection{Complete Results Of Task 2}
\label{Appendix:task2}
Table~\ref{tab:task2} presents the experimental results of Task 2, where Win Rate reflects the success rate between documents that satisfy different numbers of conditions under a multi-condition query (query10, which contains ten conditions), i.e., $d_k$ vs. $d_{k-1}$.

\begin{table*}[!t]
  \centering
  \resizebox{=\textwidth}{!}{
    \begin{tabular}{cccccccccccc}
    \toprule
Model & $d_{1}$\_vs\_$d_{0}$ & $d_{2}$\_vs\_$d_{1}$ & $d_{3}$\_vs\_$d_{2}$ & $d_{4}$\_vs\_$d_{3}$ & $d_{5}$\_vs\_$d_{4}$ & $d_{6}$\_vs\_$d_{5}$ & $d_{7}$\_vs\_$d_{6}$ & $d_{8}$\_vs\_$d_{7}$ & $d_{9}$\_vs\_$d_{8}$ & $d_{10}$\_vs\_$d_{9}$ & Avg.\\
    \midrule
    \multicolumn{11}{c}{Sparse Retriever} \\
    \midrule
    \rowcolor[rgb]{ 1,  .953,  .949} BM25  & 13.91 & 16.50  & 16.81 & 18.14 & 22.10  & 29.04 & 37.78 & 38.87 & 39.93 & 40.19 & 25.90\\
    \midrule
    \multicolumn{11}{c}{Dense Retriever} \\
    \midrule
    \rowcolor[rgb]{.949,  .949,  1} jina-embeddings-v3 & 73.43 & 70.52 & 67.45 & 66.66 & 65.32 & 63.40  & 63.15 & 62.82 & 65.13 & 60.35 & 65.82\\
    
    \rowcolor[rgb]{.949,  .949,  1} gte-large-en-v1.5 & 76.74 & 73.85 & 72.70  & 69.91 & 70.32 & 68.05 & 67.39 & 64.09 & 65.14 & 62.58 & 69.08\\
    
    \rowcolor[rgb]{.949,  .949,  1} NV-Embed-v2 & 82.57 & 76.39 & 74.45 & 72.10  & 73.27 & 69.15 & 69.48 & 66.74 & 68.75 & 71.57 &72.45 \\
    
    \rowcolor[rgb]{.949,  .949,  1} bge-en-icl & 79.40  & 70.58 & 69.36 & 68.13 & 64.80  & 63.12 & 63.01 & 61.72 & 61.31 & 63.69 & 66.51\\
    
    \rowcolor[rgb]{.949,  .949,  1} gte-Qwen2-7B-instruct & 79.84 & 74.02 & 70.57 & 69.97 & 65.44 & 60.54 & 61.35 & 59.42 & 59.55 & 60.40 & 66.11\\
    
    \rowcolor[rgb]{.949,  .949,  1} gte-Qwen2-1.5B-instruct & 74.30  & 71.80  & 72.28 & 68.49 & 69.32 & 65.69 & 67.08 & 64.97 & 63.46 & 65.09 & 68.25\\
    
    \rowcolor[rgb]{.949,  .949,  1} e5-mistral-7b-instruct & 75.11 & 67.88 & 62.73 & 58.61 & 56.87 & 54.52 & 55.26 & 54.03 & 56.68 & 61.94 & 60.36\\
    
    \rowcolor[rgb]{.949,  .949,  1} GritLM-7B & 79.59 & 77.73 & 73.40  & 74.71 & 75.56 & 72.15 & 73.52 & 71.87 & 72.01 & 75.21 & 74.58\\
    
    \rowcolor[rgb]{.949,  .949,  1} LLM2Vec & 83.50  & 74.25 & 73.43 & 72.24 & 70.36 & 67.21 & 66.99 & 67.07 & 66.49 & 67.48 & 70.90\\
    \midrule
    \multicolumn{11}{c}{Fine-tuned Reranker} \\
    \midrule
    \rowcolor[rgb]{  .949,  1,  .949} bge-reranker-v2-m3 & 76.08 & 68.06 & 63.83 & 62.06 & 60.65 & 58.35 & 54.79 & 50.60  & 48.57 & 44.96 & 58.80\\
    
    \rowcolor[rgb]{ .949,  1,  .949} bge-reranker-v2-gemma & 87.98 & 82.08 & 78.07 & 77.10  & 76.21 & 72.13 & 68.84 & 65.63 & 62.32 & 56.13 & 72.65\\
    
    \rowcolor[rgb]{  .949,  1,  .949} followIR & 61.99 & 59.82 & 60.87 & 60.76 & 59.91 & 56.41 & 51.63 & 47.93 & 44.71 & 43.52 & 54.76\\
    \rowcolor[rgb]{  .949,  1,  .949} RankZephyr & 90.20 & 86.04 & 83.96 & 83.14 & 82.23 & 79.41 & 76.07 & 72.43 & 70.26 & 66.84 & 79.06\\
    \midrule 
    \multicolumn{12}{c}{Zero-shot LLM for Ranking} \\
    \midrule
    \rowcolor[rgb]{  .949,  .949,  .949}  GPT-4o & 93.37 & 92.76 & 92.20 & 91.16 & 90.51 & 89.55 & 88.38 & 86.82 & 85.96 & 85.26 & 89.60\\
    
    \bottomrule
    \end{tabular}%
    }
  \caption{Average Win Rate Comparison Between Documents in Task 2}
  \label{tab:task2}%
\end{table*}%

\subsection{Complete Results Of Document Length}
\label{Appendix:length}
Table~\ref{tab:512} and Table~\ref{tab:1024} present the effect of document length on retrieval performance, with documents padded to 512 and 1024 words, respectively. We use repeated filler text, such as \textit{“The grass is green. The sky is blue. The sun is yellow. Here we go. There and back again.”}, following the setting in \citet{wang2023improving}.
The filler text is inserted between the original document sentences until the total text length reaches 512 or 1024 words.

\begin{table*}[htbp]
  \centering
  \resizebox{=\textwidth}{!}{
    \begin{tabular}{cccccccccccc}
    \toprule
Model & $d_{1}$\_vs\_$d_{0}$ & $d_{2}$\_vs\_$d_{1}$ & $d_{3}$\_vs\_$d_{2}$ & $d_{4}$\_vs\_$d_{3}$ & $d_{5}$\_vs\_$d_{4}$ & $d_{6}$\_vs\_$d_{5}$ & $d_{7}$\_vs\_$d_{6}$ & $d_{8}$\_vs\_$d_{7}$ & $d_{9}$\_vs\_$d_{8}$ & $d^+$\_vs\_$d_{9}$ & Avg.\\
    \midrule
    \multicolumn{12}{c}{Sparse Retriever} \\
    \midrule
    \rowcolor[rgb]{ 1,  .953,  .949} BM25  & 11.91 & 14.34 & 14.75 & 14.45 & 15.99 & 24.75 & 36.58 & 37.68 & 38.02 & 39.34 & 24.78 \\
    \midrule
    \multicolumn{12}{c}{Dense Retriever} \\
    \midrule
    \rowcolor[rgb]{ .949,  .949,  1} jina-embeddings-v3 & 64.8  & 60.11 & 58.61 & 58.31 & 57.83 & 56.36 & 56.42 & 57.20  & 58.56 & 58.59 & 58.68 \\

    \rowcolor[rgb]{ .949,  .949,  1} gte-large-en-v1.5 & 66.63 & 61.15 & 57.66 & 59.44 & 56.26 & 55.10  & 53.64 & 53.14 & 51.34 & 54.51 & 56.89 \\

    \rowcolor[rgb]{ .949,  .949,  1} NV-Embed-v2 & 68.83 & 62.87 & 60.97 & 62.16 & 61.78 & 61.80  & 62.04 & 61.10  & 62.75 & 64.78 & 62.91 \\

    \rowcolor[rgb]{ .949,  .949,  1} bge-en-icl & 70.55 & 62.18 & 59.35 & 60.33 & 59.70  & 60.19 & 59.28 & 58.95 & 60.16 & 60.23 & 61.09 \\

    \rowcolor[rgb]{ .949,  .949,  1} gte-Qwen2-7B-instruct & 68.63 & 64.87 & 61.91 & 60.28 & 61.98 & 58.70  & 59.28 & 59.06 & 60.28 & 59.69 & 61.47 \\

    \rowcolor[rgb]{ .949,  .949,  1} gte-Qwen2-1.5B-instruct & 69.51 & 66.38 & 63.47 & 61.89 & 60.13 & 58.76 & 58.57 & 57.36 & 58.62 & 62.11 & 61.68 \\

    \rowcolor[rgb]{ .949,  .949,  1} e5-mistral-7b-instruct & 69.98 & 63.50  & 60.30  & 59.3  & 56.77 & 54.52 & 53.05 & 53.14 & 51.47 & 53.99 & 57.60 \\

    \rowcolor[rgb]{ .949,  .949,  1} GritLM-7B & 73.46 & 70.37 & 69.19 & 70.36 & 68.36 & 67.05 & 68.34 & 61.55 & 58.38 & 59.57 & 66.66 \\

    \rowcolor[rgb]{ .949,  .949,  1} LLM2Vec & 70.57 & 68.37 & 67.10  & 67.10  & 66.35 & 58.73 & 36.04 & 36.81 & 35.37 & 37.03 & 54.35 \\
    \midrule
    \multicolumn{12}{c}{Fine-tuned Reranker} \\
    \midrule
    \rowcolor[rgb]{ .949,  1,  .949} bge-reranker-v2-m3 & 73.65 & 66.55 & 63.28 & 61.65 & 54.60  & 38.54 & 28.42 & 27.92 & 28.14 & 28.02 & 47.08 \\

    \rowcolor[rgb]{ .949,  1,  .949} bge-reranker-v2-gemma & 82.39 & 77.63 & 71.80  & 70.00    & 65.10  & 56.91 & 41.75 & 32.01 & 28.84 & 29.96 & 55.64 \\

    \rowcolor[rgb]{ .949,  1,  .949} followIR & 50.48 & 51.78 & 49.64 & 46.27 & 37.60  & 25.22 & 24.41 & 24.51 & 25.67 & 24.71 & 36.03 \\
    \rowcolor[rgb]{  .949,  1,  .949} RankZephyr & 85.41 & 79.91 & 72.91 & 70.41 & 64.41 & 52.91 & 47.91 & 45.41 & 42.30 & 42.56 & 60.41\\
    \midrule 
    \multicolumn{12}{c}{Zero-shot LLM for Ranking} \\
    \midrule
    \rowcolor[rgb]{  .949,  .949,  .949} GPT-4o & 89.41 & 88.15 & 87.50 & 87.80 & 86.30 & 84.90 & 83.20 & 82.95 & 81.10 & 80.35 & 85.17\\
    \bottomrule
    \end{tabular}%
    }
    \caption{\small Effect of document length on retrieval performance (padded to 512 words).}
  \label{tab:512}%
\end{table*}%

\begin{table*}[htbp]
  \centering
  \resizebox{=\textwidth}{!}{
    \begin{tabular}{cccccccccccc}
    \toprule
Model & $d_{1}$\_vs\_$d_{0}$ & $d_{2}$\_vs\_$d_{1}$ & $d_{3}$\_vs\_$d_{2}$ & $d_{4}$\_vs\_$d_{3}$ & $d_{5}$\_vs\_$d_{4}$ & $d_{6}$\_vs\_$d_{5}$ & $d_{7}$\_vs\_$d_{6}$ & $d_{8}$\_vs\_$d_{7}$ & $d_{9}$\_vs\_$d_{8}$ & $d^+$\_vs\_$d_{9}$ & Avg.\\
    \midrule
    \multicolumn{12}{c}{Sparse Retriever} \\
    \midrule
    \rowcolor[rgb]{ 1,  .953,  .949} BM25  & 12.25  & 14.32  & 14.86  & 14.81  & 15.97  & 24.48  & 36.00  & 37.53  & 38.85  & 39.53  & 24.86  \\
    \midrule
    \multicolumn{12}{c}{Dense Retriever} \\
    \midrule
    \rowcolor[rgb]{ .949,  .949,  1} jina-embeddings-v3 & 64.56  & 59.74  & 58.13  & 57.20  & 55.47  & 55.90  & 55.33  & 53.70  & 54.75  & 54.81  & 56.96  \\

    \rowcolor[rgb]{ .949,  .949,  1} gte-large-en-v1.5 & 68.62  & 61.61  & 58.47  & 54.77  & 54.97  & 54.52  & 54.44  & 49.88  & 39.09  & 38.34  & 53.47  \\

    \rowcolor[rgb]{ .949,  .949,  1} NV-Embed-v2 & 59.23  & 61.26  & 62.58  & 62.81  & 64.57  & 63.95  & 62.98  & 63.49  & 65.65  & 67.55  & 63.41  \\

    \rowcolor[rgb]{ .949,  .949,  1} bge-en-icl & 66.08  & 61.93  & 60.83  & 59.63  & 59.34  & 61.04  & 61.08  & 51.67  & 36.00  & 35.95  & 55.36  \\

    \rowcolor[rgb]{ .949,  .949,  1} gte-Qwen2-7B-instruct & 66.06  & 63.23  & 63.36  & 61.35  & 59.63  & 58.56  & 56.62  & 57.97  & 36.61  & 35.96  & 55.94  \\

    \rowcolor[rgb]{ .949,  .949,  1} gte-Qwen2-1.5B-instruct & 68.46  & 63.66  & 62.02  & 62.21  & 59.47  & 60.78  & 59.38  & 58.71  & 35.01  & 34.91  & 56.46  \\

    \rowcolor[rgb]{ .949,  .949,  1} e5-mistral-7b-instruct & 66.97  & 61.11  & 59.05  & 54.53  & 54.40  & 54.47  & 53.02  & 54.55  & 53.88  & 53.57  & 56.56  \\

    \rowcolor[rgb]{ .949,  .949,  1} GritLM-7B & 71.47  & 67.97  & 69.41  & 56.02  & 54.49  & 52.58  & 54.21  & 56.35  & 54.87  & 57.26  & 59.46  \\

    \rowcolor[rgb]{ .949,  .949,  1} LLM2Vec & 74.81  & 72.05  & 71.55  & 26.85  & 26.68  & 26.35  & 25.24  & 26.16  & 26.04  & 27.27  & 40.30  \\
    \midrule
    \multicolumn{12}{c}{Fine-tuned Reranker} \\
    \midrule
    \rowcolor[rgb]{ .949,  1,  .949} bge-reranker-v2-m3 & 76.56  & 70.75  & 56.83  & 18.33  & 19.86  & 19.09  & 18.55  & 20.75  & 19.88  & 21.78  & 34.24  \\

    \rowcolor[rgb]{ .949,  1,  .949} bge-reranker-v2-gemma & 83.48  & 77.02  & 66.34  & 19.76  & 20.54  & 19.81  & 20.64  & 19.34  & 20.65  & 21.46  & 36.90  \\

    \rowcolor[rgb]{ .949,  1,  .949} followIR & 52.36  & 51.43  & 19.70  & 18.35  & 17.15  & 17.11  & 17.78  & 16.94  & 17.77  & 18.75  & 24.73  \\
    \rowcolor[rgb]{  .949,  1,  .949} RankZephyr & 84.34 & 76.84 & 66.64 & 35.14 & 33.34 & 34.04 & 31.54 & 32.64 & 30.74 & 31.24 & 45.65\\
    \midrule 
    \multicolumn{12}{c}{Zero-shot LLM for Ranking} \\
    \midrule
    \rowcolor[rgb]{  .949,  .949,  .949} GPT-4o & 88.80 & 86.50 & 85.80 & 86.10 & 83.50 & 82.00 & 80.20 & 79.50 & 78.30 & 77.63 & 82.83\\
    \bottomrule
    \end{tabular}%
    }
    \caption{Effect of document length on retrieval performance (padded to 1024 words).}
  \label{tab:1024}%
\end{table*}%

\subsection{Comparative Results for other LLMs}
We evaluated two additional LLMs—Qwen3-32B and Gemini-2.0-Flash—on the same MultiConIR Task 1. As shown in Table \ref{tab:llms}, the results indicate that Gemini-2.0-Flash attains the highest WinRate (\textbf{91.20}), marginally outperforming GPT-4o (90.47), with Qwen3-32B in third place (86.61). GPT-4o is therefore not significantly superior, confirming that MultiConIR does not inherently favor GPT-4o. 
\begin{table}
\centering
\resizebox{=0.95\textwidth}{!}{
\begin{tabular}{lccccccccccc}
\toprule
\textbf{Model} & Q1 & Q2 & Q3 & Q4 & Q5 & Q6 & Q7 & Q8 & Q9 & Q10 & \textbf{Avg.} \\
\midrule
Qwen3-32B        & 91.20 & 90.64 & 89.55 & 88.08 & 86.91 & 85.35 & 84.74 & 84.44 & 83.27 & 81.88 & 86.61 \\
GPT-4o           & 95.49 & 94.89 & 93.71 & 92.11 & 90.81 & 89.08 & 88.43 & 88.08 & 86.82 & 85.26 & 90.47 \\
Gemini-2.0-flash & 95.67 & 95.10 & 94.06 & 92.63 & 91.50 & 89.99 & 89.38 & 89.08 & 87.95 & 86.61 & \textbf{91.20} \\
\bottomrule
\end{tabular}
}
\caption{Performance comparison across Q1--Q10 queries and average score.}
\label{tab:llms}
\end{table}

% \subsection{Attention heatmap of cross-encoder model}
% \label{sec:attention heatmap}
% Figure \ref{fig:cross-heatmap} shows the attention-score heat map produced by the cross-encoder reranker (bge-reranker-m3) for the input query–document pair. As we progressively add conditions to the query and compute the attention distribution between each condition and its corresponding segment in the document, we observe that the cross-encoder allocates attention unevenly across positions.
% \begin{figure}
%     \centering
%     \includegraphics[width=1\linewidth]{cross_heatmap.png}
%     \caption{\small Attention heatmap of cross-encoder model (bge-reranker-m3).}
%     \label{fig:cross-heatmap}
% \end{figure}
% Fig.\ref{fig:cross-heatmap} 

\section{Findings In Constructing \textbf{\textsc{Multiconir}} Dataset}
\label{Appendix:findings}
\subsection{The Use Of LLM-generated Data in Retrieval}
\label{Appendix:ai data}

In recent years, artificial datasets generated by LLMs have become a common practice for training and evaluating retrieval models~\citep{weller2024followir,shi-etal-2025-personax,Shi2024,zhang2025featuresdeserveattentiongraphguided,han2025attributestextualgenesleveraging}. For instance, E5-Mistral~\citep{wang-etal-2024-improving-text} rely entirely on LLM-generated datasets for fine-tuning. While this approach can significantly expand training corpora, prior studies have highlighted its potential drawbacks, including introducing inherit linguistic biases of the underlying LLMs~\citep{shumailov2024ai}, potentially constraining the retrieval model’s performance and generalizability. Furthermore, purely artificial data often lacks the contextual richness and complexity found in real-world retrieval scenarios~\citep{li2023synthetic, wang2024generateddatahelpcontrastive}, making it difficult to capture the actual needs of users’ queries accurately.

During our dataset construction, we observed similar issues. When using LLM-generated transformations to modify positive documents into hard negatives, the model often restructured expressions to fit its learned patterns, even when explicitly instructed to modify only a few condition-related words while keeping the rest unchanged.
For example, in the legal documents dataset, a positive sentence like:
\textit{“The defendant was convicted of fraud under Section 420 of the Penal Code and sentenced to five years in prison.”}
was frequently modified by the LLM into a generic pattern, such as:
\textit{“The defendant was found guilty of fraud and received a prison sentence.”}

Similarly, in medical case documents, a sentence like:
\textit{“The patient reported experiencing persistent chest pain and shortness of breath, leading to a diagnosis of angina.”}
was often transformed into a standardized version:
\textit{“The patient was diagnosed with a heart condition after reporting chest pain.”}

These modifications eroded the diversity and long-tail characteristics of real-world data, reducing the fine-grained variability necessary for retrieval tasks. Instead of preserving rich domain-specific details, LLM-generated transformations tended to normalize distinct cases into overly generic patterns, which could misrepresent real-world retrieval challenges.

Empirical results further confirm the limitations of fully LLM-generated training data. The E5-Mistral model, which relies entirely on synthetic data, performs the worst on \textsc{MultiConIR}. In Task 1, as shown in Table \ref{tab:task1}, it exhibits the highest performance decline (16.93\%) among retrieval models, and in Task 2, as shown in Table \ref{tab:task2}, its average win rate (60.36\%) is the lowest among retrieval models, trailing the second-worst model (Jina-Embeddings-V2) by 5\%. These results reinforce the generalization challenges posed by fully synthetic datasets in retrieval tasks, highlighting the importance of incorporating real-world document structures and constraints in training data.

To mitigate this, our pipeline minimizes document-wide modifications, instead restricting LLM interventions to condition sentences only. This targeted approach preserves real-world data authenticity while introducing controlled semantic perturbations, ensuring that retrieval models are trained on meaningful and realistic hard negatives rather than fully synthetic documents.

\subsection{Impact of Different Hard Negative Construction Strategies}
\label{Appendix:different HN}

To systematically examine the impact of hard negative sentence (HNS) construction on retrieval models, we experimented with two distinct approaches:
(1) Key Information Modification – altering critical details while maintaining overall sentence structure (applied to books, movies, medical cases, and legal documents).
(2) Keyword Retention with Dummy Information – keeping all original keywords intact while injecting irrelevant dummy information (used for the people dataset).

A key objective of this study was to investigate how different HNS construction strategies affect retrieval difficulty. Our initial hypothesis was that the second approach (retaining keywords but adding dummy information) would pose a greater challenge for retrieval models, particularly Dense Retrievers, since hard negatives in this setting contain all the key terms present in positive documents.

\begin{figure*}[!ht]
    \centering
    \begin{subfigure}{0.49\textwidth}
        \centering
        \includegraphics[width=\linewidth]{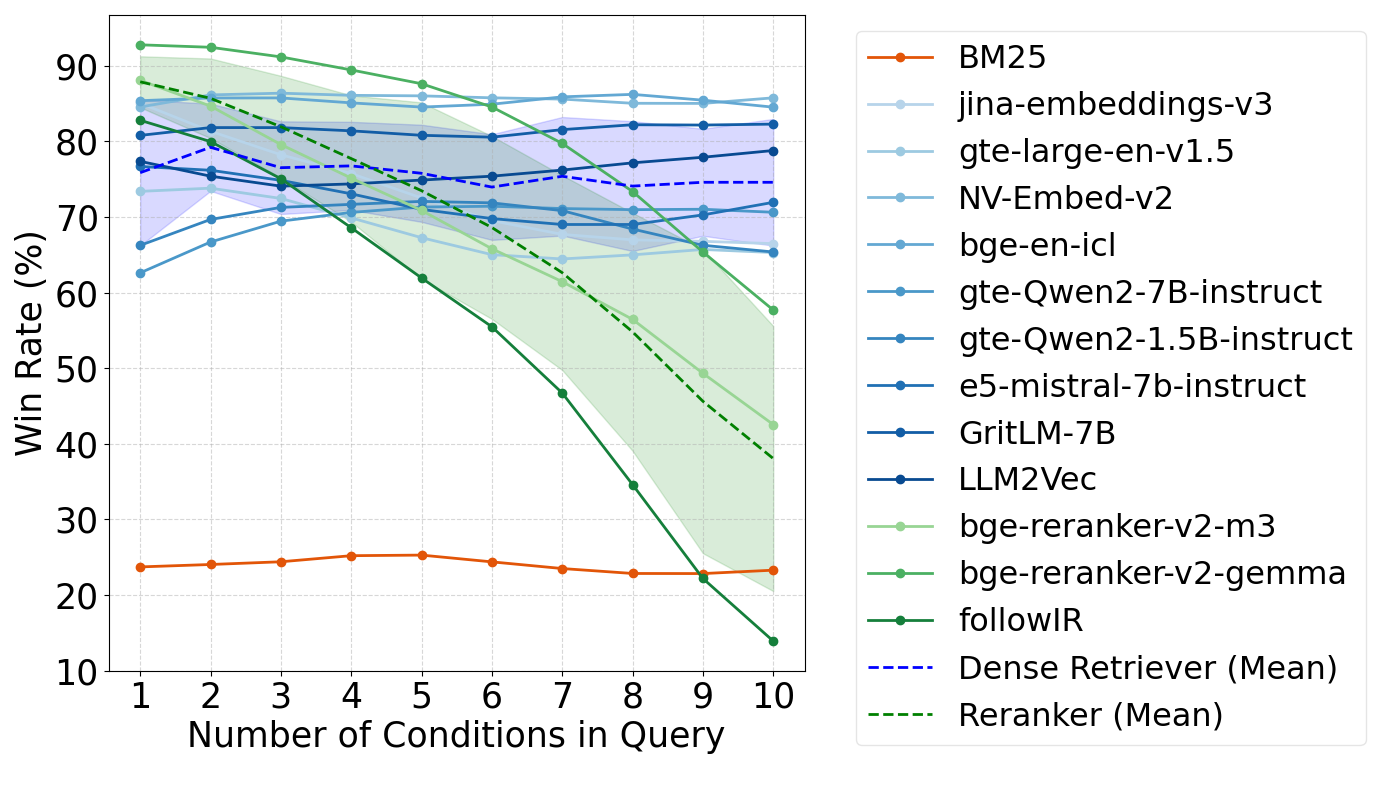}
        \caption{\small Task1 on Peole dataset}
        \label{fig:people}
    \end{subfigure}
    \hfill
    \begin{subfigure}{0.49\textwidth}
        \centering
        \includegraphics[width=\linewidth]{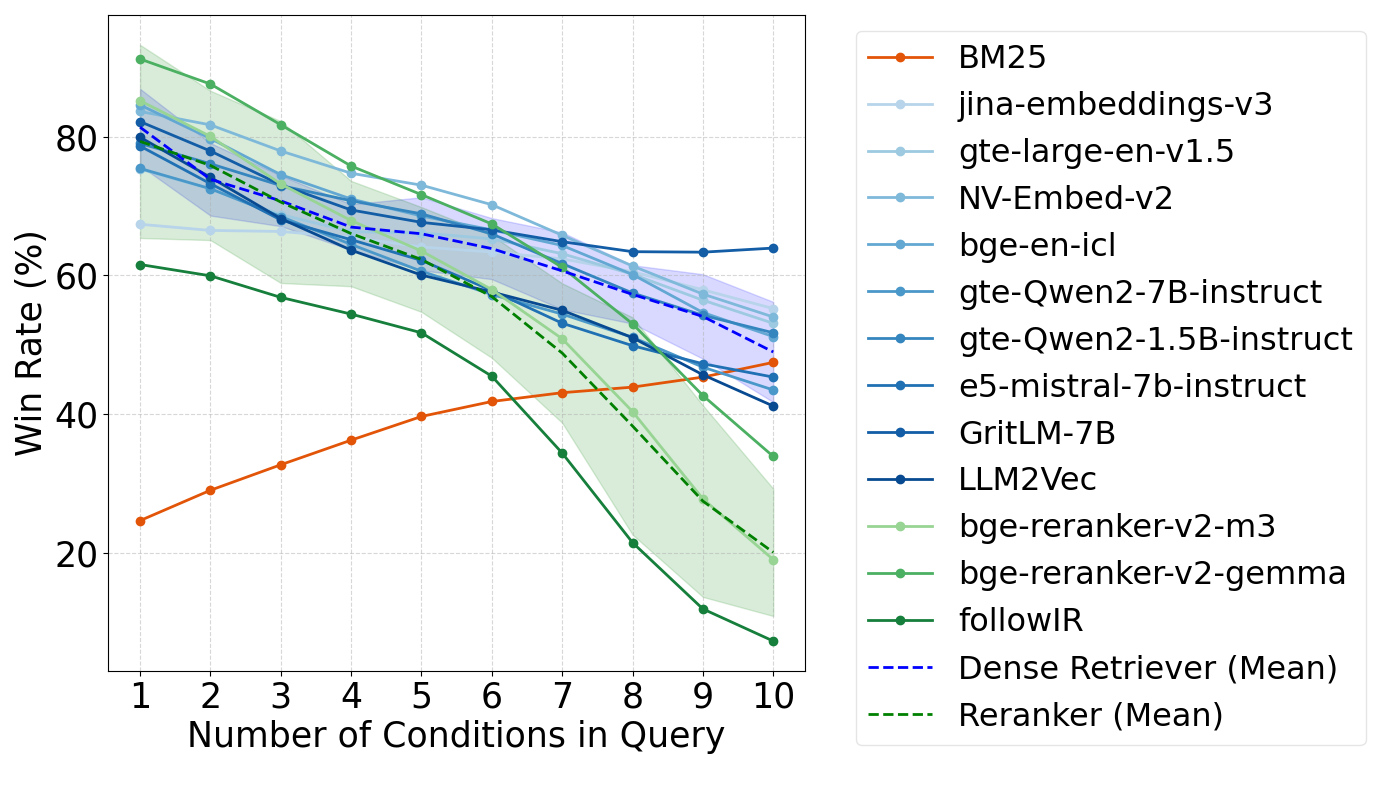}
        \caption{\small Task1 on Legal dataset}
        \label{fig:legal}
    \end{subfigure}
    \caption{Impact of different HNS construction strategies.}
    \label{fig:people and legal}
\end{figure*}

However, our experimental results contradicted this expectation. As shown in Fig.\ref{fig:people and legal} on the people dataset, Dense Retrieval models remained highly stable, demonstrating a strong ability to differentiate semantic nuances even when all keywords were retained. This suggests that Dense Retrieval primarily relies on contextual embeddings rather than simple keyword matching, allowing it to distinguish between truly relevant documents and distractors with superficial lexical overlap.

In contrast, Reranker models exhibited a significant performance drop when dealing with dummy-information-based HNS. This suggests that Rerankers are more sensitive to this type of negative construction, likely due to their cross-encoder or generative architectures, which process both the query and document jointly. Since Rerankers typically assign scores based on fine-grained textual relevance, the presence of keyword overlap without genuine semantic alignment may mislead them more than Dense Retrieval models.

These findings highlight important considerations for hard negative sampling in multi-condition retrieval. While Dense Retrievers appear robust to surface-level keyword retention, Rerankers are more vulnerable to semantically misleading negatives, suggesting that future retrieval pipelines should adapt negative sampling strategies based on the target retrieval model architecture. 

\end{document}